\documentclass[fleqn,usenatbib]{mnras}
\usepackage{natbib}
\usepackage{amsmath}
\usepackage{ulem}
\usepackage{adjustbox}
\usepackage[version=3]{mhchem}
\usepackage{lscape}
\usepackage{hyperref}
\usepackage{deluxetable}
\usepackage{caption}
\usepackage{graphicx}
\usepackage[displaymath, mathlines]{lineno}
\usepackage[title]{appendix}
\usepackage{xcolor} 

\linenumbers

\title[Radio Observations of V745 Sco]{The Symbiotic Recurrent Nova V745 Sco at Radio Wavelengths
}
\author[I. Molina et al.]{Isabella Molina$^{1}$\thanks{E-mail: molinai1@msu.edu},
Laura Chomiuk$^{1}$,
Justin D.\ Linford$^{2}$,
Elias Aydi$^{1}$,
Amy J.\ Mioduszewski$^{2}$,
\newauthor
Koji Mukai$^{3,4}$,
Kirill V.\ Sokolovsky$^{5}$,
Jay Strader$^{1}$,
Peter Craig$^{1}$,
Dillon Dong$^{2}$,
Chelsea E.\ Harris$^1$,
\newauthor
Miriam M.\ Nyamai$^{6}$,
Michael P.\ Rupen$^{7}$,
Jennifer L.\ Sokoloski$^{8}$,
Frederick M.\ Walter$^{9}$,
\newauthor
Jennifer H.~S.\ Weston$^{10}$, and
Montana N.\ Williams$^{11}$
\\
$^{1}$Center for Data Intensive and Time Domain Astronomy, Department of Physics and Astronomy, Michigan State University,\\
East Lansing, MI 48824, USA\\
$^{2}$National Radio Astronomy Observatory, P.O. Box O, Socorro, NM 87801, USA\\
$^{3}$Center for Space Science and Technology, University of Maryland Baltimore County, Baltimore, MD 21250, USA\\
$^{4}$CRESST and X-ray Astrophysics Laboratory, NASA/GSFC, Greenbelt MD 20771 USA\\
$^{5}$Department of Astronomy, University of Illinois at Urbana-Champaign, 1002 W. Green Street, Urbana, IL 61801 USA\\
$^{6}$South African Radio Observatory (SARAO), Liesbeek House, River Park Liesbeek Parkway, Settlers Way, Mowbray, Cape Town, 7705\\
$^{7}$Herzberg Institute of Astrophysics, National Research Council of Canada, Penticton, BC V2A 6J9, Canada\\
$^{8}$Columbia Astrophysics Laboratory, Columbia University, New York, NY, USA\\
$^{9}$Department of Physics and Astronomy, Stony Brook University, Stony Brook, New York 11794, USA\\
$^{10}$US Naval Observatory, 3450 Massachusetts Ave NW, Washington, DC 20392\\
$^{11}$ Department of Physics, New Mexico Tech, 801 Leroy Pl., Socorro, NM 87801, USA
}

\begin{document}
\nolinenumbers
\label{firstpage}
\pagerange{\pageref{firstpage}--\pageref{lastpage}}
\maketitle
  
\begin{abstract}
V745 Sco is a Galactic symbiotic recurrent nova with nova eruptions in 1937, 1989 and 2014. We study the behaviour of V745 Sco at radio wavelengths (0.6--37\,GHz), covering both its 1989 and 2014 eruptions and informed by optical, X-ray, and $\gamma$-ray data. 
The radio light curves are synchrotron-dominated. Surprisingly, compared to expectations for synchrotron emission from explosive transients such as radio supernovae, the
light curves spanning 0.6--37 GHz all peak around the same time ($\sim$18--26 days after eruption) and with similar flux densities (5--9 mJy). 
We model the synchrotron light curves as interaction of the nova ejecta with the red giant wind, 
but find that simple spherically symmetric models with wind-like circumstellar material (CSM) cannot explain the radio light curve. Instead, we conclude that the shock suddenly breaks out of a dense CSM absorbing screen around 20 days after eruption, and then expands into a relatively low density wind ($\dot{M}_{out} \approx 10^{-9}-10^{-8}$ M$_{\odot}$ yr$^{-1}$ for $v_w = 10$ km s$^{-1}$) out to $\sim$1 year post-eruption. The dense, close-in CSM may be an equatorial density enhancement or a more spherical red giant wind with  $\dot{M}_{in} \approx [5-10] \times 10^{-7}$ M$_{\odot}$ yr$^{-1}$, truncated beyond several $\times 10^{14}$ cm.
The outer lower-density CSM would not be visible in typical radio observations of Type Ia supernovae: V745 Sco cannot be ruled out as a Type Ia progenitor based on CSM constraints alone.
Complementary constraints from the free--free radio optical depth and the synchrotron luminosity imply the shock is efficient at accelerating relativistic electrons and amplifying magnetic fields, with $\epsilon_e$ and $\epsilon_B \approx 0.01-0.1$.
\end{abstract}


\section{Introduction} \label{intro}
A nova is a thermonuclear explosion that ignites at the bottom of a layer of accreted material on a white dwarf (WD) in a binary system \citep{Gallagher&Starrfield78, Bode&Evans08, Chomiuk+21}. The companion star is usually a main-sequence star, but is occasionally a more evolved sub-giant or giant star. 
The companion transfers H-rich gas onto the WD, accumulating an envelope of accreted material on its surface. As this material is compressed, the pressure and temperature at the base of the accreted layer increase, and nuclear reactions accelerate, until conditions are reached for thermonuclear runaway \citep{Starrfield+16}. This leads to a sudden increase in luminosity, and the accreted layer is expelled from the WD at velocities of $\sim 500-5000$ km s$^{-1}$. Some material from the WD may be mixed into the accreted envelope and also expelled in the eruption \citep[see][and references therein]{2011Natur.478..490C,2018ApJ...860..110S,2020MNRAS.497.2569S}. The binary survives this nova eruption, and the accretion process can continue, implying that multiple nova eruptions may recur over time in a given binary. The properties of nova eruptions, including recurrence time and ejecta mass ($\sim 10^{-7}-10^{-3}$ M$_{\odot}$), are thought to primarily depend on the WD mass and accretion rate \citep{Townsley&Bildsten04,Yaron+05, Hillman+2016}. While most novae have recurrence times significantly longer than human time-scales, if the WD is massive enough and the accretion occurs at a fast enough rate, then the recurrence time between nova eruptions can shorten to decades, years, or even months \citep{Yaron+05, Wolf+13, 2016ApJ...833..149D}. Systems where multiple nova eruptions have been observed are called `recurrent novae', with 10 recurrent novae confirmed in the Milky Way \citep{Schaefer10} and a similar number of recurrent novae candidates \citep{Pagnotta&Schaefer14}. Four of these ten are `symbiotic' binaries with giant companions \citep{Kenyon86}, implying that evolved companion stars may be over-represented amongst recurrent novae (compared to classical novae with longer recurrence times; e.g. \citealt{Ozdonmez+18}; although we note that the symbiotic fraction remains poorly constrained amongst novae; see e.g. \citealt{Williams+16} for a pioneering study). The prevalence of giant companions amongst recurrent novae is often explained by  accretion fed by red giant winds reaching higher rates than accretion fed by a dwarf companion filling its Roche Lobe \citep{Chen+16, Kemp+21}.

\subsection{Recurrent Novae as Potential SN Ia Progenitors}
Recurrent novae are potential candidate progenitors of Type Ia supernovae (SNe Ia). A SN Ia is a type of supernova that takes place in a binary system with a carbon-oxygen WD and a companion star \citep{Maoz+14, Liu+23}.  In contrast with a nova, when a SN Ia occurs, the WD is completely destroyed. The companion star may be either an evolved star (the single degenerate channel; \citealt{Whelan_Iben73}) or another WD (the double degenerate channel; \citealt{Webbink84}).  In the single-degenerate scenario, the companion star transfers material onto the WD, causing it to grow in mass and density. Electron degeneracy pressure prevents the WD from collapsing due to gravity until the WD  approaches Chandrasekhar mass, when it becomes unstable and collapses, producing a SN Ia. In the double-degenerate class of scenarios,  the binary system consists of two WDs, at least one of which must be composed of carbon-oxygen, but neither of which needs to grow near the Chandrasekhar mass. The dominant channel producing most SNe Ia remains an open question, but observational tests imply that both single and double degenerate scenarios likely produce SNe Ia of various subtypes \citep[e.g.][]{2023MNRAS.521.1162D, 2023Natur.617..477K, 2023ApJ...952...24H, 2023ApJ...956L..34S}.

Classical novae are challenging to the single degenerate scenario for SNe Ia because they make it very difficult for the WD to retain the mass it has accreted, and may actually  lead to the WD shrinking in mass \citep{Gehrz+98,2015arXiv150202665S}. However, models predict that recurrent novae tend to eject less accreted material,  allowing the WD to grow over time \citep{Yaron+05} potentially near Chandrasekhar mass. The WDs in most recurrent novae have been observed to be massive, approaching the Chandrasekhar mass. Using the effective temperature of the WD, V745 Sco was found to have a $M_{WD} > 1.3 M_{\odot}$ (\citealt{Page+15}; see \citet{Wolf+13} for the relationship between WD effective temperature and mass). The effective temperatures of the WDs in two other recurrent novae, RS Oph and V3890 Sgr also suggest high masses. V3890 Sgr has a $M_{WD} = 1.25 - 1.3M_{\odot}$ and RS Oph was found to have a $M_{WD} = 1.2M_{\odot}$  \citep{Page+20,Osborne+11b}.

While these results support massive WDs as the hosts of recurrent novae, it  is an open question if these massive WDs are composed of carbon-oxygen or oxygen-neon (a ONe WD will likely collapse into a neutron star rather than exploding as a SN Ia; \citealt{Gutierrez+96,2018MNRAS.481..439W,2019MNRAS.484..698R}). Still, recurrent novae with evolved companions remain compelling candidates for SN Ia progenitors---and are also some of the most testable candidate progenitors. For symbiotic recurrent novae with giant companions, the giant drives a dense wind, polluting the circumbinary environment. Red giant mass loss rates range from $\dot{M} \approx 10^{-8} - 10^{-6}\ M_{\odot}\, yr^{-1} $ \citep{Seaquist&Taylor90} with velocities of the order, $10-100$ km s$^{-1}$ \citep{Gehrz+71, Kudritzki+78}. Recurrent nova eruptions can then sweep up and shape this wind into a series of shells and cavities \citep{Wood-Vasey&Sokoloski06}, potentially leading to parsec-scale nova super-remnants \citep{Darnley+19,Shara+24}. This dense, structured circumstellar material (CSM) is detectable, both shaping how a symbiotic recurrent nova appears and also leaving observable signatures in SNe Ia, should these systems be SN Ia progenitors. The SN~Ia--CSM interaction might be detectable as radio synchrotron emission \citep{Panagia+06, Chomiuk+16, Lundqvist+20},  X-ray emission \citep{Russell&Immler12, Sand+21}, optical absorption lines \citep{Patat+07, Maguire+13}, or even as features in the optical light curve \citep{2023MNRAS.522.6035M}. By observing nearby recurrent novae and measuring the density and distribution of their CSM, we can obtain a real-world benchmark for this SN Ia progenitor scenario, which can in turn be compared with observations of SNe Ia. A goal of this paper is to make progress on these questions with a detailed study of one Galactic symbiotic recurrent nova, V745 Sco.

\subsection{V745 Sco}
V745 Sco is a symbiotic recurrent nova with eruptions observed on May 10, 1937 \citep{Plaut58}, July 30.08, 1989 \citep{Liller89}, and February 6.69, 2014 \citep{Waagen14}. The companion star was determined to be an M6 III giant based on TiO bands \citep{Duerbeck_89} or an M4 III giant using CO absorption features in infrared spectra \citep{Harrison+93}. \citet{Schaefer10} used photometry to estimate the orbital period to be around 510 days. However, \citet{Mroz+14} show with higher cadence, long-term photometry from OGLE \citep{Udalski03} that a period of 510 days (or 255 days) is not significant, and instead the optical light curve of V745 Sco during quiescence is dominated by pulsations of the giant component. These semi-regular pulsations of the red giant have periods of 136.5 and 77.4 days. 

\begin{figure}
    \centering \includegraphics[width=\columnwidth]{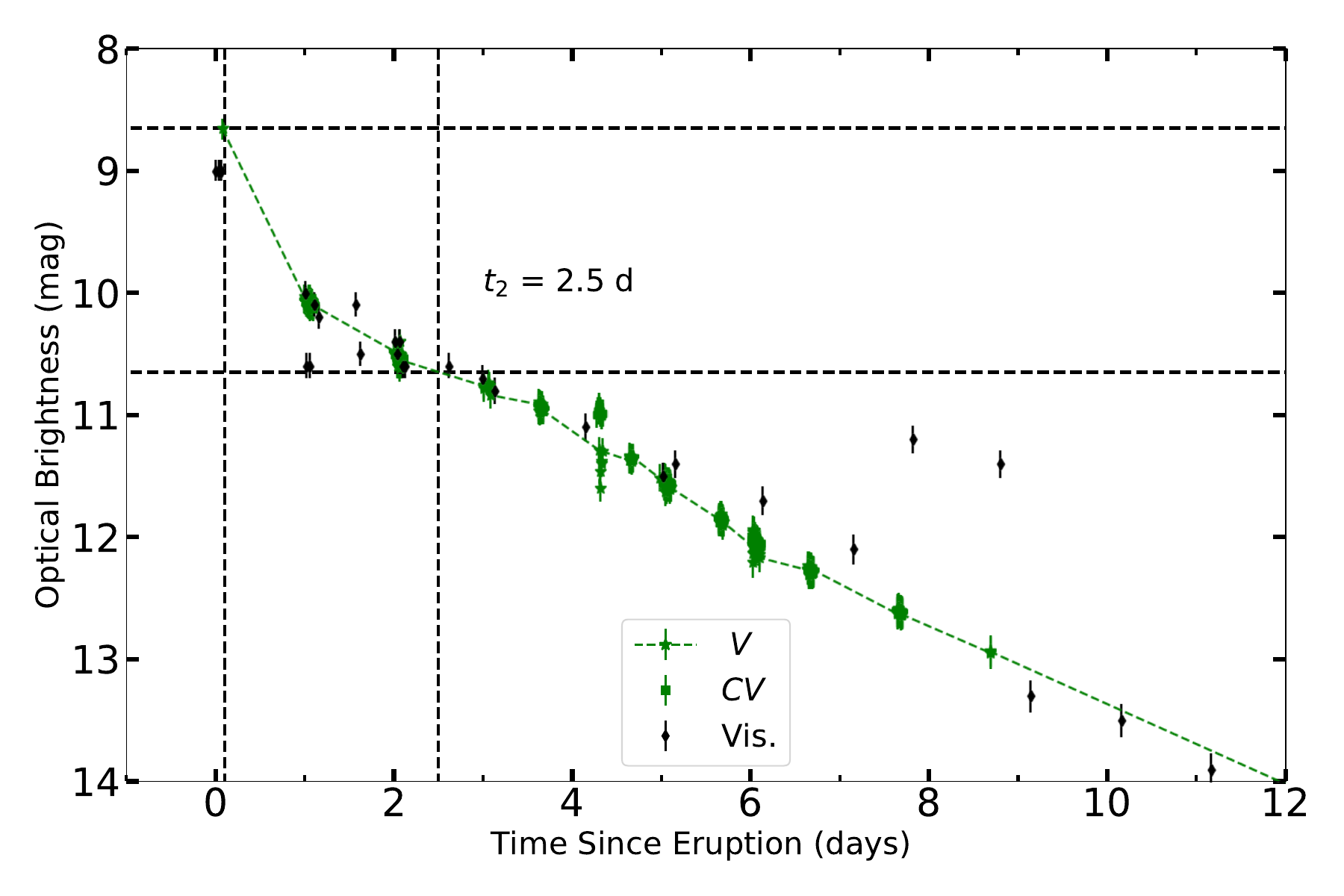}
    \caption{The AAVSO optical light curve of V745~Sco during the first 12 days of the 2014 eruption ($t_0 =$ 2014 Feb 6.69). The vertical dashed lines represent $t_{\mathrm{peak}}$ and $t_2$. The horizontal dashed lines represent $V_{\mathrm{peak}}$ and $V_{\mathrm{peak}}-2$.}
    \label{Fig:optical_LC}
\end{figure}

V745 Sco erupts with a relatively short recurrence time, with only 25 years between the 1989 and 2014 eruptions. 
The optical light curve of V745 Sco's 2014 eruption is presented in Figure~\ref{Fig:optical_LC}, created using  publicly available data reported to the American Association of Variable Stars (AAVSO; \citealt{AAVSODATA}). The data consist of $V$-band photometry and visual estimates. The nova reached optical peak at $V_{\rm peak} = 8.66$\,mag on 2014 Feb 06.77, a few hours after discovery (2014 Feb 06.69, which we take as $t_0$). The optical light curve shows a rapid evolution with the time for the nova to decline two magnitudes from optical peak, $t_2 =$ 2.5\,days, making it a very fast nova. The 1989 $V$-band light curve starts at $V=9.7$ mag and declined slower than the 2014 light curve with a $t_2 =5$ days  \citep{Schaefer10,Sekiguchi+90}, perhaps indicating that the peak of the light curve was not observed. The rapid optical decline of V745 Sco's eruptions increases the likelihood that observers may have missed one eruption or more. 

The 2014 eruption was detected at X-ray, UV, optical, near-IR and radio wavelengths \citep{Page+15, Orio+15, Drake+16, Delgado&Hernanz19, Mroz+14, Banerjee+14, Kantharia+16}. In addition, marginal detections of GeV $\gamma$-rays were obtained with {\it Fermi}/LAT on 2014 Feb 6 and 7, at 2.4 and 2.5$\sigma$ level significance \citep{Franckowiak+18}. {\it Swift} observed the nova soon after eruption, obtaining X-ray light curves with XRT as well as UV coverage with UVOT \citep{Page+15}. 
Single-epoch X-ray observations were obtained with {\it Chandra} grating spectrometers and {\it NuSTAR}, and showed that the temperature of the shock-heated gas was decreasing over time \citep{Orio+15, Drake+16}. This drop in temperature was interpreted as the shock wave decelerating as the expanding ejecta sweep up CSM \citep{Orio+15, Drake+16, Orlando+17, Delgado&Hernanz19}. 
\citet{Drake+16} and \citet{Orlando+17} determined that the shocks producing the observed X-ray emission are the result of collisions with an aspherical CSM, characterized by an equatorial density enhancement. \cite{Drake+16} estimated the mass that is lost during the eruption to be $M_{ej} = 10^{-7}$ M$_{\odot}$ (see also \citealt{Page+15}) and concluded that this is less than the ignition mass, implying that the WD may be growing in mass over the course of the accretion--eruption cycle.

\citet{Banerjee+14} analysed near IR data taken with the Mount Abu Infrared Observatory starting 1.3 days after discovery of the 2014 eruption. The Pa$\beta$ line profiles narrowed rapidly and were used to estimate the ejecta velocity via their full width at half maximum (FWHM). These FWHM values do not fit a $t^{-1/3}$ decline, as expected for the Sedov Taylor phase of nova ejecta interacting with CSM, which they claim further supports the idea that the CSM is asymmetric. 
\citet{Kantharia+16} presented low-frequency radio observations (610 MHz and 235 MHz) of V745 Sco's 2014 eruption obtained with the Giant Metrewave Radio Telescope (GMRT) and compared them to 1.4 GHz observations of the 1989 eruption.

The goal of this paper is to revisit the radio behaviour of V745 Sco over the course of its 1989 and 2014 eruptions, using newly published data of unprecedented quality from the Very Large Array (VLA; both before and after its upgrade to the Karl G.\ Jansky VLA) covering frequencies from 1--37 GHz. In \S2 we present the observations used in our analysis. In \S3 we detail the distance measurements made for V745 Sco based on high-resolution optical spectroscopy and three-dimensional dust maps. In \S4 we present the optical spectroscopic data and discuss whether they can be used to estimate the velocity and radius of the nova shock. In \S5 we present the radio behaviour of V745 Sco: the multi-frequency radio light curves, brightness temperature evolution, and spectral evolution. In \S6 we  model the radio synchrotron emission from the nova using the formalism of \citet{Chevalier82}, and discuss whether a V745 Sco-like progenitor would be detectable, given current progenitor constraints on Type Ia supernovae. We summarize our results and discuss future steps in \S7.

\section{Observations and Data Reduction}
\subsection{Optical Spectroscopy of the 2014 Eruption}\label{sec:smarts}
We revisit the spectroscopic evolution of V745~Sco using publicly available optical spectra obtained using the Small and Moderate Aperture Research Telescope System (SMARTS) 1.5m telescope and its CHIRON optical spectrograph \citep{Walter+12}. The observations covered the wavelength range of $4080 - 8900$ \textup{\AA} with a resolution  of $R=27,800$. Observations began February 9, 2014 (3 days after eruption) and ended May 13, 2014 (96 days after eruption), and were obtained with near-daily cadence.  In Table~\ref{tab:spectral} we present dates of the observations, along with the measured line width values as well as the integrated flux of the H$\alpha$ line (see \S \ref{sec:blastwave} for more details). 

\subsection{Radio Observations of the 1989 Eruption}\label{sec:1989radiodata}
Observations with the VLA started on September 1.9 1989 (about 34 days after eruption) and ended February 15 1990 (200 days after eruption). Observations were obtained under VLA program codes AL202 (PI W.\ Lewin), AH383, AH389, and AH390 (all three with PI R.\ Hjellming). The observations were obtained before the 2010 Expanded VLA upgrade, and were conducted in full-Stokes `continuum mode' with two closely spaced spectral windows each providing 43 MHz of bandwidth. The frequency bands observed were L band (1.50 GHz), C band (4.86 GHz), X band (8.44 GHz), and U band (14.94 GHz). 3C286 was used as the absolute flux density calibrator, and $1733-130$ and $1743-038$ as complex gain calibrators.

We edited, calibrated and imaged these data using standard routines in \textsc{AIPS} \citep{Greisen03}. The data from 1990 Feb 2 were obtained during `move time' between D and A configurations, and without an absolute flux calibrator. For this epoch, we took the flux density of the complex gain calibrator $1733-130$ from the subsequent epoch, and restricted the $uv$ range in imaging, but the data quality remained poor and the nova was not detected.

\begin{figure}
    \centering
    \includegraphics[width=80mm]{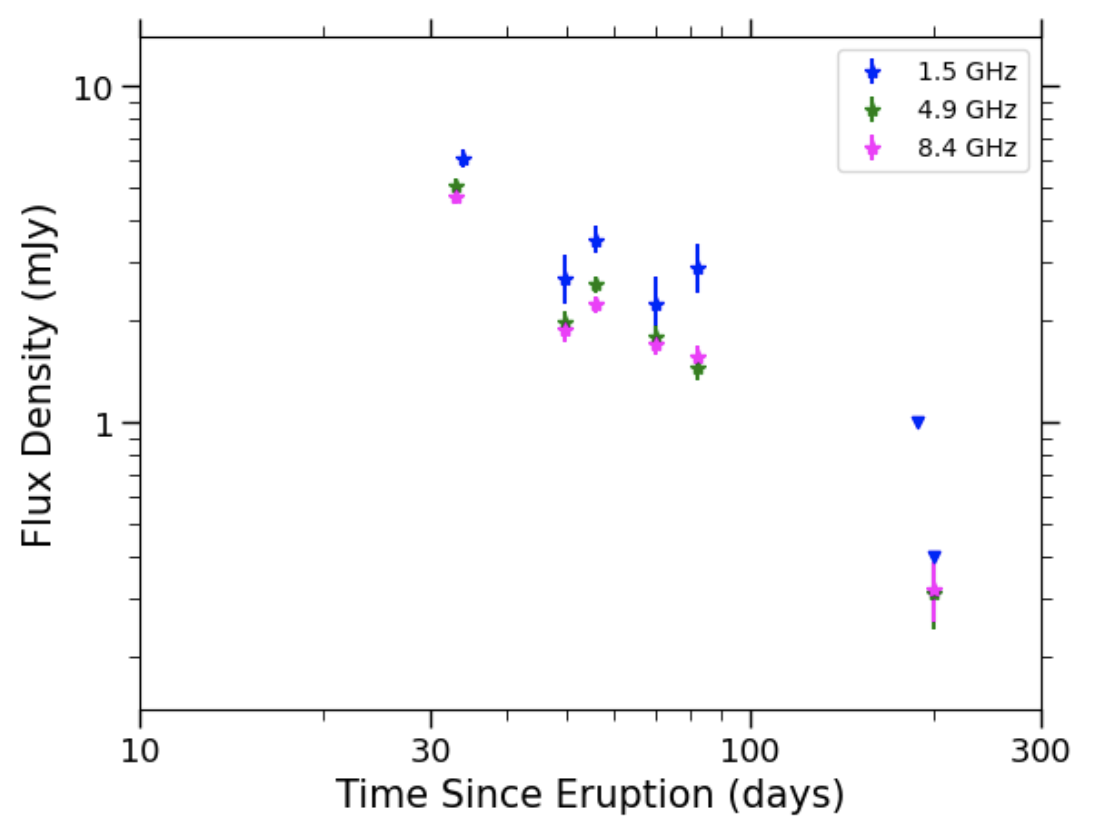}
    \caption{Multi-frequency radio light curve of V745 Sco's 1989 eruption. We took the day of discovery, July 30.08 1989, as $t_0$. Non detections are shown as downward-facing triangles.}
    \label{fig:1989lc}
\end{figure}

Flux densities were measured by fitting a Gaussian with width fixed to that of the synthesized beam, and we use a 3$\sigma$ detection threshold (upper limits for non-detected epochs are $3\sigma$ significance).  Radio flux densities are listed in Table \ref{tab:1989} and the multi-frequency radio light curve is plotted in Figure~\ref{fig:1989lc}. For lower frequency observations ($<$10 GHz), we assumed a calibration error of 5~per~cent, while for higher frequency observations ($>$10 GHz) we assumed 10~per~cent calibration errors.  These calibration errors were not considered when determining which measurements were detections (significant at $>3 \sigma$ level), but are quoted in Table \ref{tab:1989} as part of the uncertainty in flux density measurements. 

We note that the 1.5 GHz data were presented in \citet{Kantharia+16}, but the data presented here represent an independent reduction. Our measurements generally agree with theirs within the uncertainties, with the exception of the 1989 Oct 20 epoch, where our flux measurement is more than twice as bright as \citet{Kantharia+16}. This discrepancy can be easily understood, because one of the spectral windows was not fringing on the calibrator and target on this day; we flagged this spectral window and only used the one good window, but if we had not, the resulting flux would have been erroneously low. For this reason, we prefer the value listed in Table  \ref{tab:1989}. This is the first time the higher frequency data of the 1989 eruption have been published. 

During the last observation on day 200, the VLA was in its extended A configuration, which yields the highest angular resolution. While Galactic novae are occasionally resolved in this configuration (e.g. QU Vul- \citealt{Taylor+88}; V959 Mon- \citealt{Chomiuk+14}), V745 Sco had already faded significantly and a resolved image could not be produced from the data.

\begin{deluxetable}{lcccc}
\centering
\tablewidth{0 pt}
\tablecaption{Radio Observations of V745 Sco's 1989 Eruption}
\label{tab:1989}
\tablehead{UT Date& $t-t_0$ & Freq &  Flux $\pm$ Error & Config\\
 & (days) & (GHz) & (mJy) & } 
\startdata
1989 Sep 1.9& 34 & 1.49 & $6.13 \pm 0.38$ & CnB \\ 
1989 Sep 1.9& 34 & 4.86 & $5.06 \pm 0.27$ & CnB \\
1989 Sep 1.9& 34 & 8.44 & $4.69 \pm 0.24$ & CnB \\

1989 Sep 17.9& 50 & 1.49 & $2.69 \pm 0.44$ & CnB\\
1989 Sep 17.9& 50 & 4.86 & $1.98 \pm 0.17$ & CnB\\
1989 Sep 17.9& 50 & 8.44 & $1.88 \pm 0.14$ & CnB\\

1989 Sep 24.0& 56 & 1.49 & $3.50 \pm 0.32$ & CnB\\ 
1989 Sep 24.0& 56 & 4.86 & $2.58 \pm 0.14$ & CnB\\ 
1989 Sep 24.0& 56 & 8.44 & $2.24 \pm 0.12$ & CnB\\ 

1989 Oct 7.9& 70 & 1.49 & $2.26 \pm 0.46$ & DnC\\  
1989 Oct 7.9& 70 & 4.86 & $1.79 \pm 0.15$ & DnC\\  
1989 Oct 7.9& 70 & 8.44 & $1.70 \pm 0.12$ & DnC\\  

1989 Oct 20.0& 82 & 1.51 & $2.90 \pm 0.48$ & DnC\\ 
1989 Oct 20.0& 82 & 4.86 & $1.45 \pm 0.12$ & DnC\\ 
1989 Oct 20.0& 82 & 8.44 & $1.57 \pm 0.11$ & DnC\\ 

1990 Feb 2.7 & 187 & 1.49 & $<1.00$  &D$\rightarrow$A\\  

1990 Feb 15.5 &200& 1.49 & $<0.40$ & A\\  
1990 Feb 15.5 &200& 4.86 & $ 0.31 \pm 0.07$ & A\\  
1990 Feb 15.5 &200 & 8.44 & $ 0.32 \pm 0.07$ & A\\  
1990 Feb 15.5 & 200& 14.94 & $<0.19$ & A  
\enddata
\tablenotetext{a}{We take the start of eruption $t_0$ to be 1989 July 30.08 UT.} 
\end{deluxetable}

\begin{figure*}
    \centering \includegraphics[width=6in]{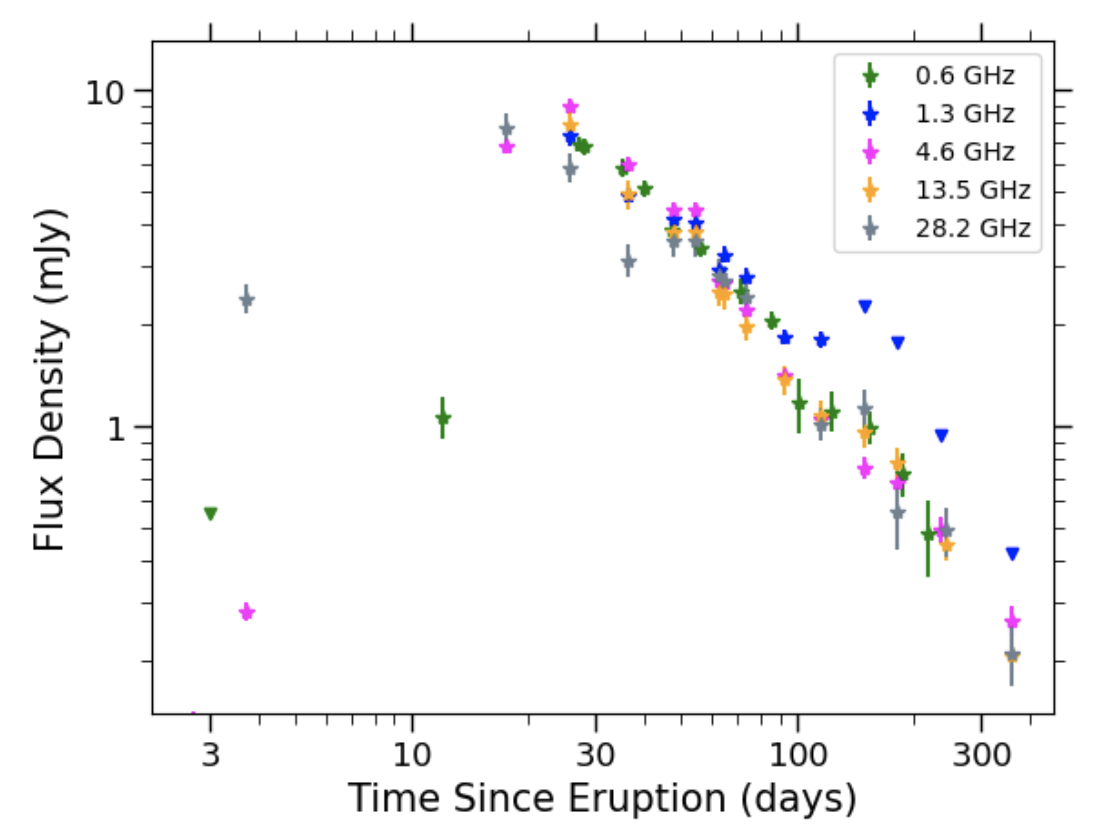}
    \caption{The multi-frequency VLA light curve of the 2014 eruption of V745 Sco, including 610 MHz data from \citet{Kantharia+16}. We take 2014 Feb 6.7 as $t_0$. Flux density measurements are plotted as stars with error bars, and 3$\sigma$ upper limits are plotted as downward-facing triangles. Note that all frequency bands appear to peak at similar flux densities and at similar times (day $\sim$18--26). Given the strong overlap in light curves at different frequencies, we only plot a representative sub-sample of frequencies here.}
    \label{fig:lcKanth2014}
\end{figure*}

\subsection{Radio Observations of the 2014 Eruption}
The upgraded VLA monitored the 2014 eruption of V745 Sco from February 8, 2014 (2.6 days after the discovery of the 2014 eruption) to February 1, 2015 (day 360) under program code 13B-057 (PI L.\ Chomiuk). The data were collected in Full Stokes continuum mode in L (1--2 GHz), C (4--8 GHz), Ku (12--18 GHz) and Ka (26.5--40 GHz) bands. The L-band observations yielded 1 GHz of bandwidth sampled with 1 MHz-wide channels, while the other bands are sampled with two independently tunable 1-GHz wide windows, providing 2 GHz of bandwidth total sampled with 2 MHz wide channels. A typical scheduling block was 1.5 hours in duration, and observations are obtained in all four receiver bands. The absolute gain calibrator 3C286 is visited for several minutes in each band, in order to calibrate the bandpass and flux density scale. Observations are made of V745 Sco in each receiver band in sequence, toggling back and forth between the nova and a complex gain calibrator (J1744$-$3116 at C, Ku, and Ka bands, and J1751$-$2524 at L band). Typical on-source times in each receiver band are 5--15 minutes. 

The VLA data were edited, calibrated, and imaged using standard routines in \textsc{AIPS}, \textsc{IRAF}, and \textsc{Difmap} \citep{Greisen03, McMullin+07, Shepherd97}. Data were imaged using a Briggs weighting with a robust value of 1. The cleaned images showed the nova as consistent with a point source in all epochs, and flux densities were measured by fitting a Gaussian with width fixed to the image synthesized beam. Flux density values are listed in Table \ref{tab:2014}, and the multi-frequency light curve is plotted in Figure~\ref{fig:lcKanth2014}. Figure~~\ref{fig:lcKanth2014} also shows the GMRT 610 MHz data overplotted \citep{Kantharia+16}. Calibration errors were set using the same procedure as explained in \S\ref{sec:1989radiodata} for the 1989 eruption.

\section{The Distance to V745 Sco}\label{sec:dist}
The distance of $7.8\pm1.8$ kpc that is commonly employed for V745 Sco is based on a flawed assumption: that the orbital period is 510 days \citep{Schaefer09, Schaefer10}. 
As stated in \S 1.2, \citet{Mroz+14} has shown that no such stable periodicity is detectable in extant photometry, calling into question the orbital period of V745 Sco. We therefore revisit the distance to V745 Sco using a different technique: combining absorption measurements from high-resolution optical spectroscopy with three dimensional dust maps of the Galaxy.

\begin{figure}
    \centering
    \includegraphics[width=\columnwidth]{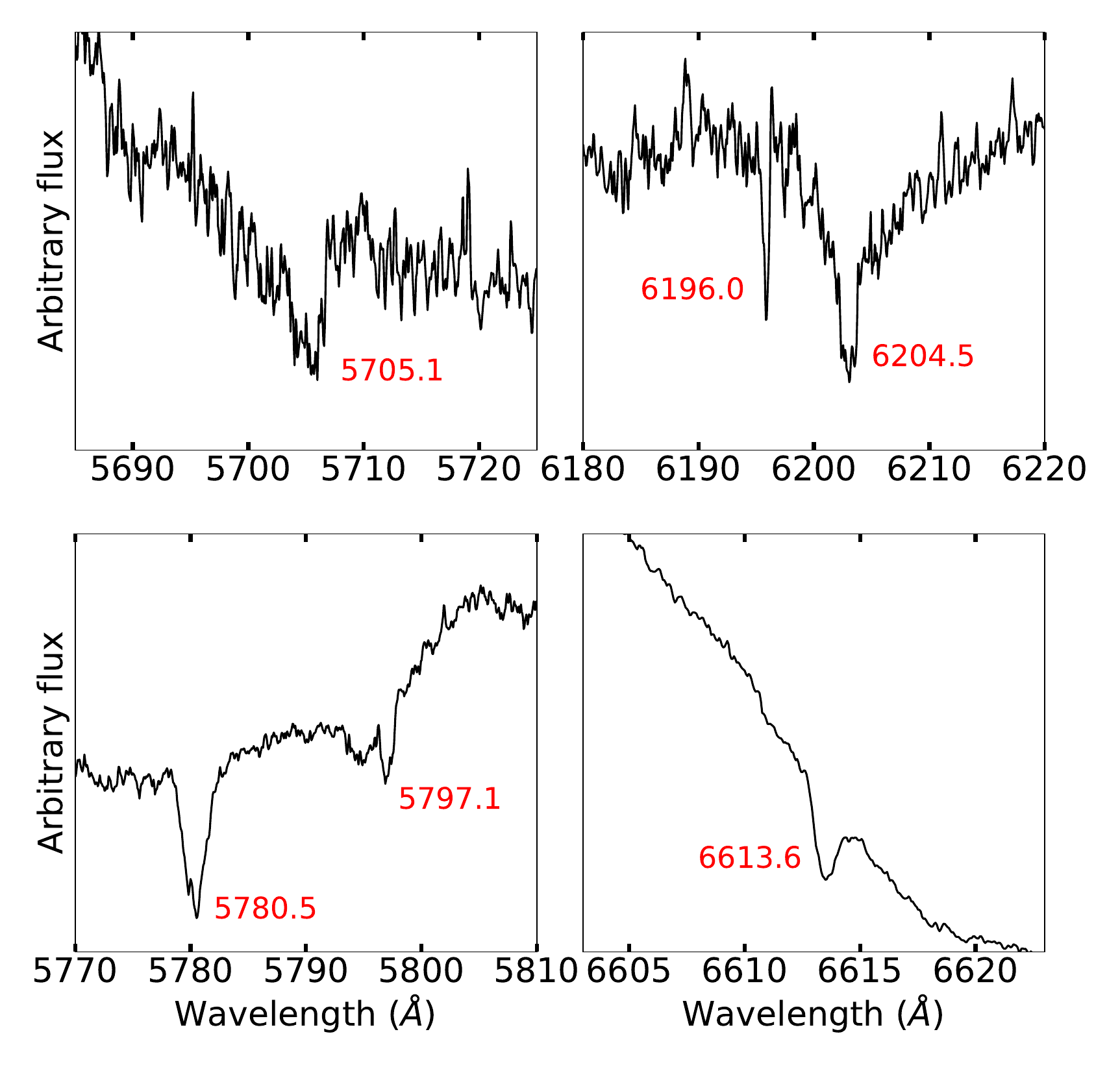}
    \caption{A sample of DIB absorption features from  used to derive the reddening towards V745~Sco. DIB absorption lines are labelled in red.}
    \label{Fig:DIBs}
\end{figure}

We use high-resolution optical spectra obtained with SMARTS/CHIRON (see \S \ref{sec:smarts}), to measure the equivalent width (EWs) of several absorption features associated with diffuse interstellar bands (DIBs), which trace dust along the line of sight. 
We focused on the DIB absorption lines at 5487.7, 5705.1, 5780.5, 5797.1, 6196.0, 6204.5, 6283.8, and 6613.6\,\AA\ (see Figure~\ref{Fig:DIBs} for a sample). We combine these measurements with the relations from \citet{Friedman_etal_2011} to derive $E(B-V)= 0.69 \pm 0.20$\,mag and $A_V = 2.14 \pm 0.6$\,mag, assuming $R_V$ = 3.1. Uncertainties are informed by the uncertainty in the EW measurements and the scatter in the relationships between EW and $E(B-V)$ from \citet{Friedman_etal_2011}, following the method in Craig et al.\ (2024, in preparation). This $A_V$ value is consistent with the total extinction along the line of sight, $A_V = 2.2$\,mag, as measured by \citealt{Schlafly&Finkbeiner11}. This corresponds to an ISM column density of $N_H = 6 \times 10^{21}$ cm$^{-2}$ \citep{Bahramian+15}.

We use the extinction derived from the DIB absorption features to constrain the distance to V745 Sco, by comparing it with the Galactic reddening maps of \citet{Chen_etal_2019}. These maps use measurements from the {\it Gaia} DR2, 2MASS, and WISE surveys, so we use the reddening laws of \citet{Wang_etal_2019} to convert $E(B-V)$ = 0.71\,mag to $E(G-K_s) = 1.56$\,mag, $E(G_B-G_R) = 0.91$\,mag, and $E(H-K_s) = 0.11$\,mag. Using these values and the maps from \citet{Chen_etal_2019} we derive a distance of at least 3.2 kpc. Beyond this distance, the extinction in the three-dimensional dust maps no longer increases, and instead flattens out to the integrated value \citep{Schlafly&Finkbeiner11}. This is not surprising, as at V745 Sco's Galactic latitude ($b = -4^{\circ}$), a distance of 3.2\,kpc implies a location 220\,pc above the plane, or $\sim$twice the scale height of dust in the Milky Way's plane \citep{Li+18}. Therefore, based off the extinction and 3D dust maps alone, we can only constrain V745 Sco's distance to $>3.2$\,kpc.

To further constrain the distance to V745 Sco, we implemented a novel strategy based on the Galactic model described by \citet[][see also \citealt{Kawash+21b}]{Kawash+22}. We generated a population of $10^7$ simulated novae in this model, distributed in proportion to stellar mass as described by a realistic  three-dimensional model of the Milky Way \citep{Robin+03}. Then for each simulated nova, we calculated its foreground $V$-band extinction using a composite three-dimensional dust model as implemented in {\tt mwdust} \citet{Bovy+16}, which combines several three-dimensional dust maps \citep{Green+19, Marshall+06, Drimmel+03}. As in \citet{Kawash+22}, we assume that the simulated novae have peak absolute magnitudes normally distributed with mean $M_V = -7.2$\,mag and standard deviation $\sigma=0.8$\,mag \citep{Shafter17}, but the absolute magnitude distribution is truncated at higher luminosities for longer $t_2$ values (not relevant for V745 Sco, where the time to decline by two magnitudes from optical peak is just $t_2 = 2.5$ days for the 2014 eruption). For each simulated nova, the simulated absolute magnitude is combined with the distance and line-of-sight extinction to infer simulated peak apparent $V$ magnitudes. We find the 300 simulated novae that are the nearest angular matches to V745 Sco. These have a median offset of 3.5\arcmin, which is lower than the typical resolution of the dust maps. Of the 300 simulated novae, we then select those that have peak $V$ magnitudes within 0.5 mag of the actual observed peak ($V_{\rm peak} = 8.66$ mag). This helps account for the uncertain measurement of $V_{peak}$ for most observed novae and allows for reasonable statistics without a more computationally intensive simulation. The resulting distance is $8.2^{+1.2}_{-1.0}$ kpc for V745 Sco, in close alignment with distances commonly used in the literature \citep[e.g.][]{Sekiguchi+90, Page+15, Delgado&Hernanz19}. We use this distance throughout the remainder of this paper.

\section{H$\alpha$ Line Profile Evolution} 
\label{sec:blastwave}
\subsection{Optical Spectral Evolution during the 2014 Eruption}
\begin{figure}
    \centering
    \includegraphics[width=\columnwidth]{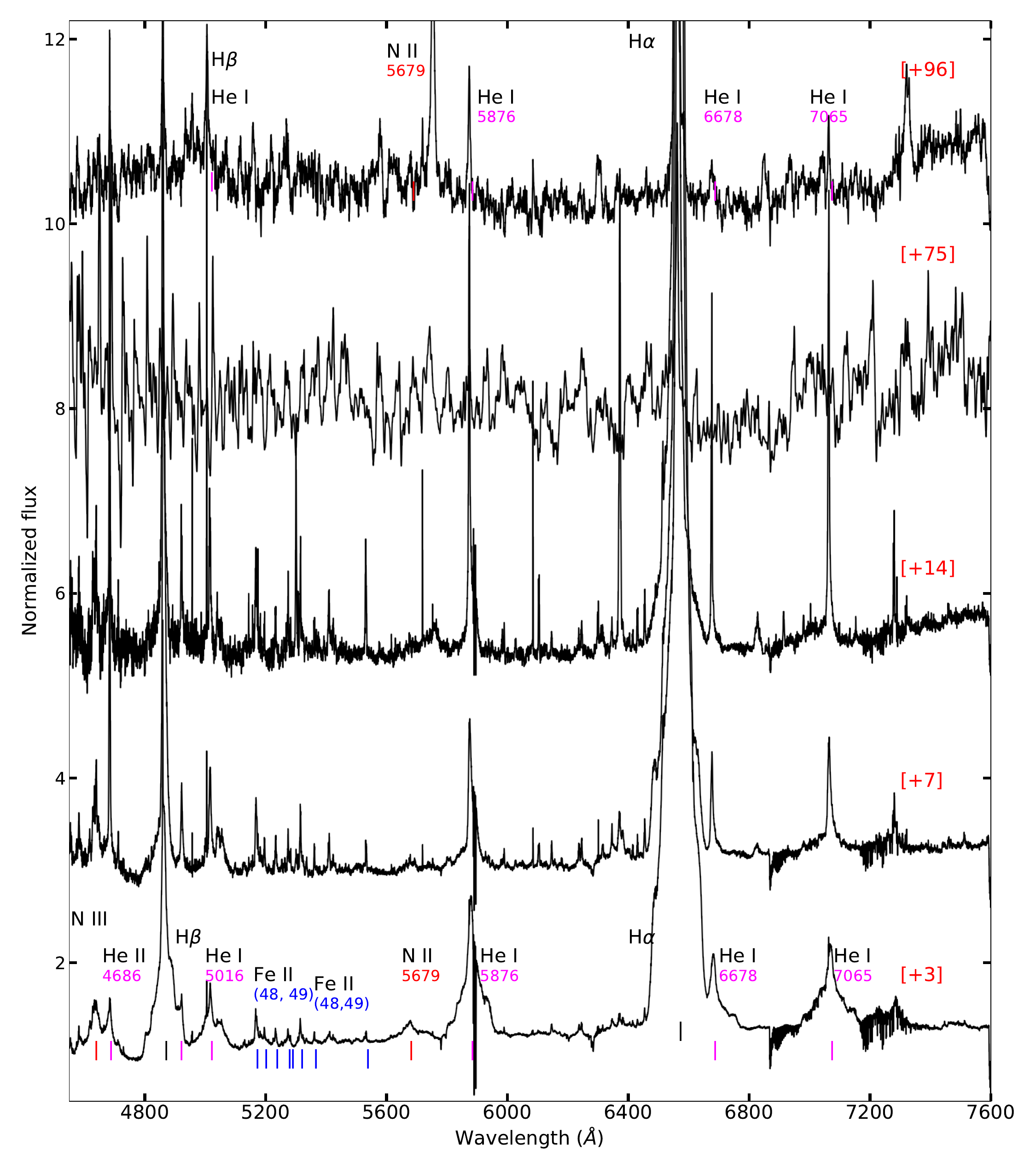}
    \caption{A sampling of optical spectra covering V745 Sco's 2014 eruption and demonstrating its spectral evolution.
    The red numbers between brackets are days after $t_0$. An offset to the normalized spectra is added for visualization purposes. Line identifications marked with vertical lines colour-coded to the element producing them are added to assist the reader.}
    \label{fig:Spec}
\end{figure}

The spectral evolution of the 2014 eruption of V745~Sco is presented in Figure~\ref{fig:Spec}, highlighting a sample of the spectral epochs covering the first $\sim$100 days of the eruption. Since the first epoch, taken 3 days into the eruption, the spectra show strong emission lines of H\,{\sc i} Balmer, He\,{\sc i}, He\,{\sc ii}, N\,{\sc iii} with weak Fe\,{\sc ii} lines per the universal spectral evolution of novae described by \cite{Aydi_etal_2023b}. The Balmer, He and N emission lines remain the dominant lines in the spectra throughout the first 100 days of the eruption. This means that the nova was first observed during a transition from the Fe\,{\sc ii} phase to the second He/N phase.

\subsection{Line Width Estimates}\label{sec:linewid}
\begin{figure}
    \centering
    \includegraphics[width=80mm]{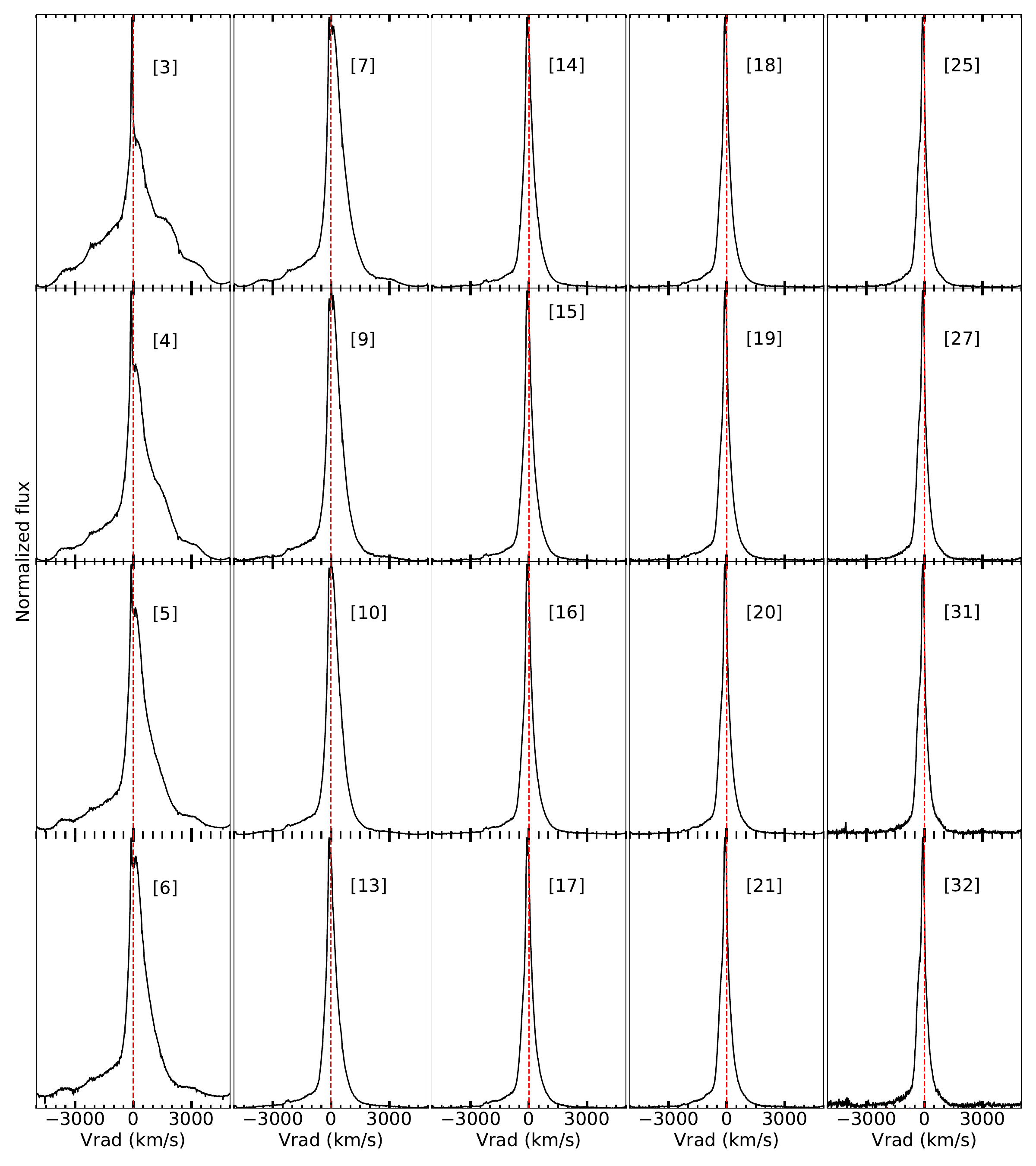}
    \caption{Evolution of the H$\alpha$ line profiles, covering 3--32 days into the 2014 eruption. In each panel, rest velocity is marked with a vertical dashed line. The line rapidly decreases in width over the first seven days. The first three epochs show multiple components in the line profile that smooth out by around day 6.}
    \label{fig:Ha}
\end{figure}

\begin{figure}
    \centering
    \includegraphics[width=80mm]{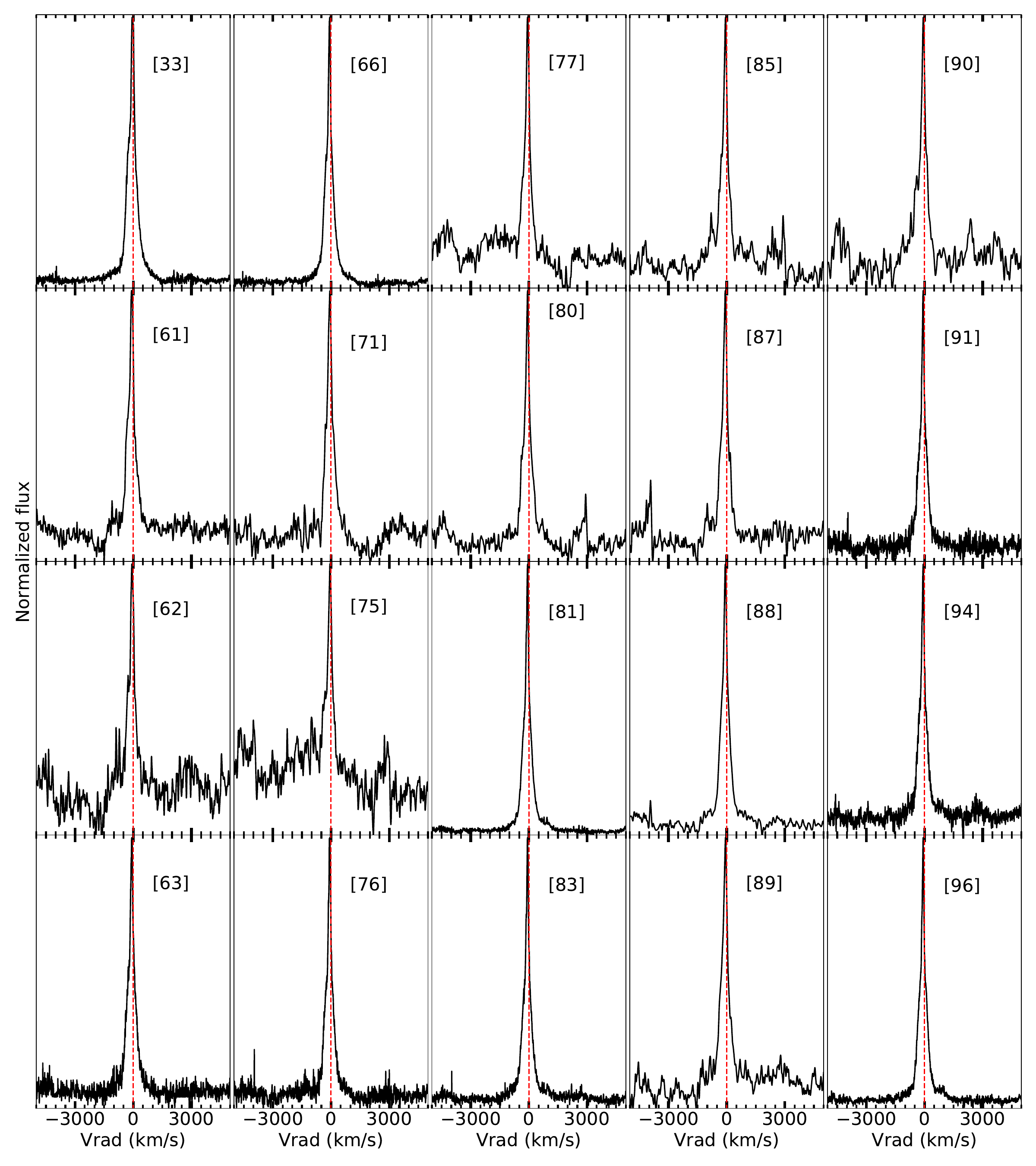}
    \caption{H$\alpha$ line profiles at later times, covering 33--96 days following the discovery of the 2014 eruption.}
    \label{fig:Ha2}
\end{figure}

The optical spectroscopic data for the 2014 eruption were used to measure the full width at zero intensity (FWZI) and FWHM of the H$\alpha$ emission line. Figures~\ref{fig:Ha} and ~\ref{fig:Ha2} show the line profile evolution of H$\alpha$ throughout the first 96 days of the eruption. 

The H$\alpha$ line starts out broad and complex; the first 15 days are the epochs that motivated \citet{Banerjee+14} to fit the line profile with a combination of a narrow and broader Gaussian, but their lower spectral resolution ($\sim$300 km s$^{-1}$ resolution) glossed over much of the observed complexity. Also visible in the earliest epochs is a bright, very narrow spike centered at $v = -134$ km s$^{-1}$ with a width of  $\sim65$ km s$^{-1}$, which has been interpreted as originating in unshocked CSM \citep{Munari+11, Banerjee+14, Zamanov+22, Azzollini+23}. 
Subsequently, the line profile narrows while the very narrow spike becomes less visible.
 
For each spectroscopic epoch, the FWZI was measured by estimating the continuum level surrounding the line and measuring the full extent of H$\alpha$ emission where it converges with the continuum. The He\,{\sc i} line at 6678\textup{~\AA} complicated this process by obscuring where the H$\alpha$ line ended on the red side. To get around this we mirrored the curve of the blue side of the H$\alpha$ line on to the red side of the curve. To measure the FWHM we used the deblending function in \textsc{IRAF} \verb|splot| to fit a Lorentzian profile to the H$\alpha$ emission line. In all epochs, the H$\alpha$ line only appears in emission with no notable absorption components or P Cygni profiles, which is unsurprising given its rapid optical decline time \citep{Aydi+20b}.  

In fitting the Lorentzian profile to the earliest epochs, we excluded the bright very narrow spike and only fit the broader underlying line. By day $\sim$10, the observed line profile has relaxed to a simpler shape reasonably well fit by a Lorentzian function.

The peak of the Lorentzian during day 10--21 was mildly but consistently lower than the observed line profile, at the $\sim$5-10~per~cent level.
By around day 60, the spectra had become significantly noisier and had to be smoothed in order to more accurately determine the continuum level. When smoothed, it was typically down to $R= 2700$, with the resulting velocity resolution less than the line width in all cases; smoothed spectra are shown in Figure~\ref{fig:Ha2}.

\begin{figure}
    \centering
    \includegraphics[width=80mm]{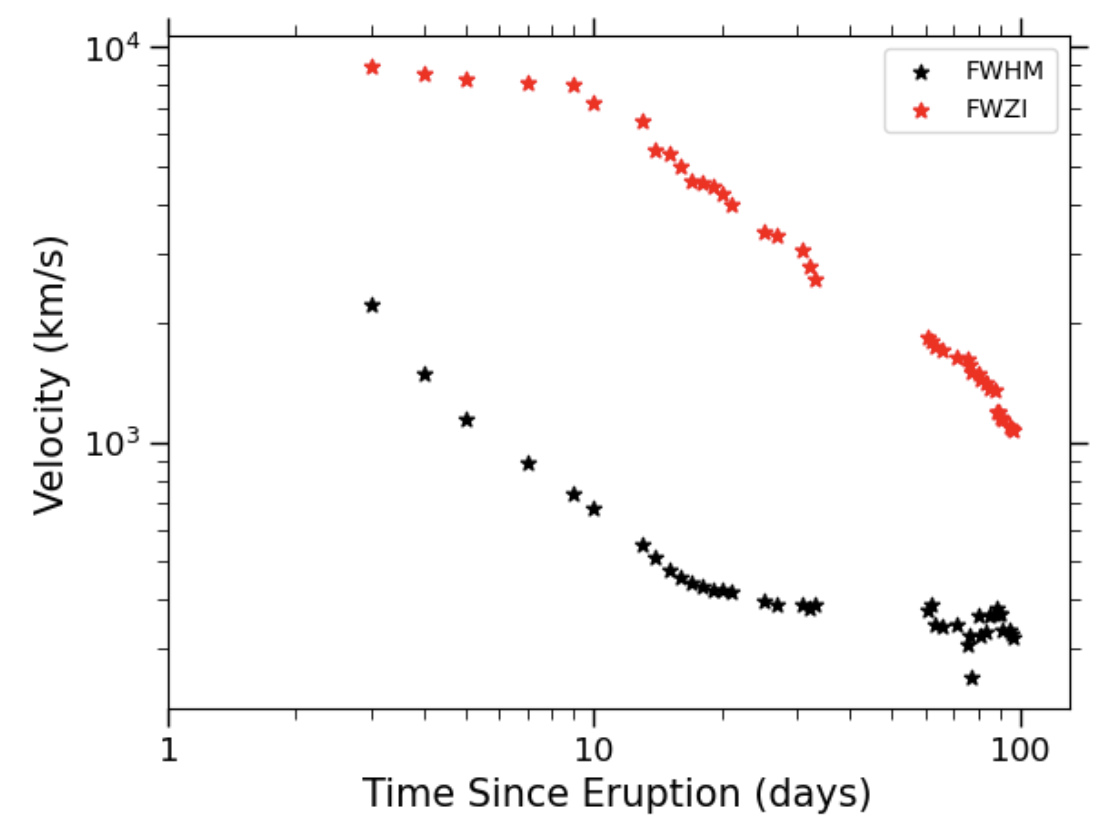}
    \caption{Width of the H$\alpha$ line (in  km $s^{-1}$) as a function of time (days since eruption) for the 2014 eruption. The line profile's FWHM is shown as black stars, and FWZI as red stars.}
    \label{fig:VFWHM}
\end{figure}

Figure~\ref{fig:VFWHM} shows how the width of the H$\alpha$ line changes over time, and measurements are tabulated in Table \ref{tab:spectral}. The line is broadest at early times, with FWZI = 8910 km s$^{-1}$ and FWHM =  2230 km s$^{-1}$ on day 3. FWZI is relatively constant at early times (FWZI $\propto t^{-0.2}$ for day $\sim3-10$), and then declines faster after  day $\sim$14 (FWZI $\propto t^{-0.8}$). The behaviour of FWHM is almost the inverse of FWZI; it declines rapidly at early times (FWHM $\propto t^{-0.9}$ for day 3--14) and then more gradually at later times (FWHM $\propto t^{-0.2}$). 
The discrepant FWHM and FWZI evolution can be explained if, in the first $\sim$10 days, the wings of the line are relatively constant in width, while the core of the line narrows. At later times, this behaviour must flip, and the core of the line has a relatively constant width, while the wings narrow. It is difficult to compare our results with those of \citet{Banerjee+14}, because they fit their line profiles with two Gaussians, but do not give the relative intensities of these dual components. 

The early narrowing of FWHM (while FWZI is nearly constant) suggests strong asymmetry, in line with the simulations of \citet{Orlando+17}. If the CSM is concentrated in an equatorial density enhancement, we expect early deceleration of ejecta in the equatorial directions, with strong H$\alpha$ emission coming from this region. Meanwhile, FWZI is measuring the largely un-decelerated polar ejecta (at least for the first $\sim$10 days or so, before the wings of the H$\alpha$ line start to fade significantly). 

We use the early line width measurements to estimate the initial kinetic energy of the ejecta by assuming an ejecta density profile, $\rho_{ej} \propto r^{-2}$, a `Hubble flow' or homologous velocity distribution $v \propto r$, and a maximum velocity of 4450 km s$^{-1}$ (matching our initial FWZI/2 measurements). We consider ejecta masses of $M_{ej} = 10^{-6}$ M$_{\odot}$ or $10^{-7}$ M$_{\odot}$ \citep{Page+15}, as expected for fast novae with short recurrence times \citep{Yaron+05}. Using the equations of \citet{Tang&Chevalier17} (see discussion in \S \ref{sec:model}), the resulting kinetic energies are $KE = 5\times10^{43}$ erg and $KE = 5\times10^{42}$ erg, respectively.

\begin{figure}
    \centering
    \includegraphics[width=80mm]{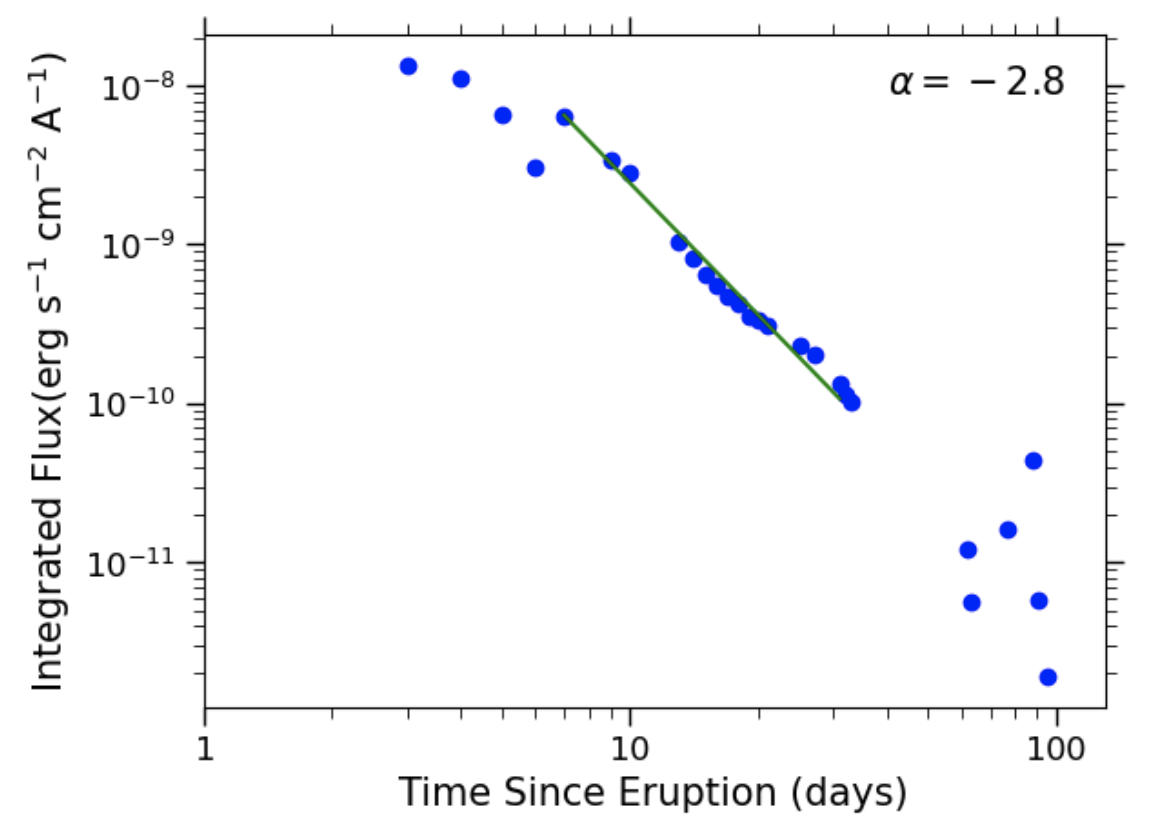}
    \caption{The integrated flux of the H$\alpha$ emission line after flux calibration as a function of time during V745 Sco's 2014 eruption. Between days 7 and 31 the flux is declining with a slope of $t^{-2.8}$.}
    \label{fig:INT}
\end{figure}

\subsection{Rapid Deceleration or Dropping Density?}\label{sec:decel}

As V745 Sco is a nova with a red giant companion, its immediate surroundings are polluted with CSM accumulated from the red giant wind \citep[e.g.][]{Seaquist&Taylor90}. When the nova ejecta expand into this CSM, mass is swept up and the ejecta decelerate, acting as a scaled-down version of a supernova remnant. 
The early evolution of the blast is characterized by a free expansion phase, when the velocity decreases slowly. This phase ends around the `Sedov time', when the blast wave has swept up a mass of CSM equivalent to the mass of the ejecta. The next phase is the Sedov-Taylor or adiabatic phase, which is characterized by a more rapid deceleration than in the free expansion phase. During this phase, as in the free expansion phase, the energy of the shock front is transferred from kinetic to internal (thermal) energy, but the total amount of energy roughly stays the same. The final phase of blast wave evolution is the radiative phase, wherein the time-scale for radiative cooling becomes shorter than the expansion time.
A revised take on the radiative phase is proposed by \citet{Delgado&Hernanz19}, wherein energy is efficiently lost from the shock by particles that are accelerated to relativistic speeds through diffusive shock acceleration and then escape the shock.

\citet{Delgado&Hernanz19} revisit the Pa$\beta$ line profiles of V745 Sco's 2014 eruption from \citet{Banerjee+14}, and note the very rapid decline in line width, $\propto t^{-0.5}$. Such a rapid deceleration cannot be explained in either the free expansion or Sedov Taylor phase (for reasonable CSM profiles, e.g. $\rho_{\rm CSM} \propto r^{-2}$; \citealt{Tang&Chevalier17}), and they conclude that the blast wave evolves from the free expansion phase straight to the radiative phase. They suggest that this rapid cooling and the quick decline in velocity is driven by very efficient particle acceleration and $\gamma$-ray radiation (see also \citealt{Tatischeff&Hernanz07}). 

Another possibility to explain the observed drastic decline in FWZI is that the ejecta and shock are dropping in density, rather than decelerating. In this case, the emission lines appear to narrow simply because their flux and signal-to-noise ratio (S/N) 
are dropping rapidly, swamping the faint wings of the lines with noise. As pointed out by  \citet{Munari+A17}, if the ejecta expand into empty space, the diameter of the ejecta ($l$) will grow with time ($l \propto t$), and the density of the ejecta will drop as $\rho \propto l^{-3} \sim t^{-3}$. The H$\alpha$ emission measure scales as $EM \propto \rho^2 l$, and the integrated flux in the H$\alpha$ line is $f_{H\alpha} \propto (l/D)^2\, EM$, where D is the distance. Hence, we expect the flux of the H$\alpha$ line to decline with time as $f_{H\alpha} \propto t^{-3}$ \citep{Munari+A17}. On the other hand, if the ejecta were sweeping up CSM, the density in the shock would drop less quickly and we would expect the decline of $f_{H\alpha}$ to be more gradual.

Novae with evolved companions show a range of behaviours; for some, the evolution of emission line profiles is consistent with a decelerating blast wave \citep{Munari+11}, while in other cases, it appears that the ejecta expand with little deceleration \citep{Munari+A17}. To test how much the H line profiles are affected by dropping density in the 2014 eruption of V745 Sco, we flux calibrated the optical spectra using contemporaneous photometry in the $R$ band from \citet{Walter+12}  and analysed the results. These photometry were obtained using the ANDICAM dual channel imager on the SMARTS 1.3m \citep{2003SPIE.4841..827D}. Figure~\ref{fig:INT} shows the resulting integrated flux of the H$\alpha$ line over time, and Table \ref{tab:spectral} lists our measurements. Between 7 to 31 days after eruption, the integrated flux declines steeply at a rate of $t^{-2.78}$. These observations of V745 Sco are consistent with $f_{H\alpha} \propto t^{-3}$, implying that it is difficult to use the H$\alpha$ emission line profile to probe blast wave dynamics.

We therefore conclude that it is likely that the apparent narrowing of the H$\alpha$ lines is primarily an observational effect driven by dropping density \citep{Munari+A17} rather than by efficient energy loss due to particle acceleration \citep{Delgado&Hernanz19}. In addition to this, the apparent continuum of the line profile is not physical. Chiron uses a single fiber and by about day 30 the sky dominates the continuum in many of the spectra. This further supports the idea that the rapid H$\alpha$ line narrowing is an observational effect. Additional reasons for preferring the `observational effect' scenario are laid out in Appendix \ref{sec:app}. This casts doubt on using the line width measurements to accurately measure the deceleration of the blast wave as a function of time in V745 Sco.  We do, however, proceed by using the initial FWZI/2 of the H$\alpha$ line, as observed on day 3, to estimate the expansion velocity of the fastest ejecta, $v = 4450$ km s$^{-1}$.

Another complication is pointed out by \citet{Banerjee+14}: instead of being a spherical wind, the CSM is likely to be denser in the orbital plane (perhaps shaped by wind Roche lobe overflow; \citealt{Mohamed&Podsiadlowski12}) and more wind-like in the polar directions. \citet{Orlando+17}  explore this complicated CSM profile with 3D hydrodynamic simulations of V745 Sco's 2014 eruption. 
They include an equatorial density enhancement in their models to create a bipolar ejecta with higher velocity in the polar directions and lower velocity in the orbital plane. They demonstrate that they can reproduce the Pa$\beta$ line width evolution observed by \citet{Banerjee+14} in their simulations without requiring efficient particle acceleration and deceleration. 
This picture implies higher shock densities while the blast is within the equatorial density enhancement, and then a drop in CSM density when the blast expands beyond the equatorial density enhancement. This would explain why the H$\alpha$ integrated flux declines at such a rapid rate around day 10: the blast is breaking out of the equatorial density enhancement and now moving in a low density region.

\section{The Behaviour of V745 Sco at Radio Wavelengths}
\subsection{Constraints on Radio Emission in Quiescence}\label{sec:quiescence}
Radio observations of symbiotic stars in quiescence can be used to estimate the density of the CSM, because the radio flux is thermal free-free emission coming from the red giant wind being ionized and heated by the accreting WD \citep{Seaquist+84, Seaquist&Taylor90}. A wind of constant mass loss rate ($\dot{M}$) and velocity ($v_w$) yields a density profile:
\begin{equation} \label{eq:winddens}
\rho_{CSM} =  \frac{\dot{M}}{4\pi v_{w}}\, r^{-2} 
\end{equation}
Assuming the radio emission is dominated by thermal free-free emission from such a wind-like CSM that is ionized and maintained at $10^4$ K, there is a one-to-one mapping between a radio flux density and $\dot{M}/v_{w}$ (Equation 2 of \citealt{Seaquist&Taylor90}).

While the VLA archive does not contain any targeted radio observations of V745 Sco out of eruption, the system was observed by the VLA Sky Survey in the S band (2--4 GHz). We downloaded three epochs of observations from the CIRADA  cutout server, obtained between 2018 and 2023. Each of these failed to detect V745 Sco down to 3$\sigma$ upper limits: $<$0.35 mJy (2018 Feb 11), $<$0.53 mJy (2020 Nov 6), and $<$0.38 mJy (2023 Jul 4).  From these radio upper limits, we estimate that the giant component of V745 Sco drives a mass-loss rate $\dot{M} < 6 \times 10^{-7}$ M$_{\odot}$ yr$^{-1}$, assuming a wind velocity of 10 km s$^{-1}$ \citep{Reimers77} and a radio spectrum where flux density $S_{\nu}$ scales with frequency $\nu$ as $S_{\nu} \propto \nu^{0.6}$, as expected for an ionized stellar wind \citep{Wright&Barlow75, Panagia&Felli75}.
 
 Our upper limit is comparable to the $\dot{M}_{\rm in}$ of the inner, denser CSM estimated by \citet{Delgado&Hernanz19}, estimated using $N_H$ values measured from X-ray spectroscopy: $\dot{M}_{\rm in} = [5-10] \times 10^{-7}$ M$_{\odot}$ yr$^{-1}$ (assuming $v_w = 10$ km s$^{-1}$). However, note that these authors argue that the inner red-giant wind is truncated beyond an outer radius of $\sim$ several $\times 10^{14}$ cm, which would decrease the expected radio emission in quiescence.

Studying a sample of symbiotic stars, \citet{Seaquist&Taylor90} measure $1 \times 10^{-8}$ M$_{\odot}$ yr$^{-1} < \dot{M} < 1 \times 10^{-6}$ M$_{\odot}$ yr$^{-1}$, so our V745 Sco upper limit fits within this range, but places V745 Sco as a symbiotic with relatively low density CSM. 

\subsection{Behaviour of the Multi frequency Radio Light Curve---and Comparison across Eruptions} \label{sec:89v14}

Radio observations of the 1989 eruption began about 34 days after discovery, on September 1, 1989 (Table \ref{tab:1989} and Figure~\ref{fig:1989lc}). The highest flux densities at all frequencies were measured on day 34, however it is very possible that the peak occurred earlier and was missed. The 1.5 GHz data has slightly higher flux density values than the higher frequencies at all epochs (take, for example, day 50 when the 1.5 GHz flux density is 0.7 mJy, or 35\% greater than the 4.9 GHz). 

The 2014 eruption is observed with better cadence and frequency coverage, and we see that the higher frequencies (28.2/36.5 GHz) appear to peak around day 17.5 (Figure~\ref{fig:lcKanth2014} and Table \ref{tab:2014}). All other frequencies peak slightly later, around 25.6 days after eruption. It is possible that the highest frequencies (e.g. 28--36 GHz) peak earlier; we note that there is a rather large observation gap between day 3.6 and 17.5. Because of this, we cannot determine the peak flux densities at these frequencies.  At frequencies ranging from 1.3 - 17.4 GHz the light curves peak at the same time and at similar flux densities, around 8 mJy. Subsequently, the radio flux declines monotonically at all frequencies.  

The multi-frequency radio light curves of both the 1989 and 2014 eruptions of V745 Sco demonstrate rather unusual behaviour when compared with other novae at radio wavelengths \citep{Chomiuk+21radio}. Classical novae (with main-sequence companions) typically show the higher frequencies peaking earlier and at greater flux densities compared to lower frequencies. This behaviour occurs if the radio luminosity is produced by thermal free-free emission and absorbed by free-free absorption, as expected for warm ($\sim 10^4$ K) ionized expanding nova ejecta \citep{Hjellming+79, Seaquist&Bode08, Finzell+18}. For V745 Sco, we note a similar time and brightness of the light curve's peak across frequencies, suggesting that free-free emission is not what is driving  the radio emission in this system. Other novae show more complex structure or multiple peaks in their radio light curves \citep{Taylor+87, Eyres+09, Weston+16a, Finzell+18, Chomiuk+21radio, Sokolovsky+23, Nyamai+23}, which is often attributed to multiple emission mechanisms (perhaps a combination of thermal and synchrotron emission, the relative contributions changing with time). 

\begin{figure}
    \centering
    \includegraphics[width=85mm]{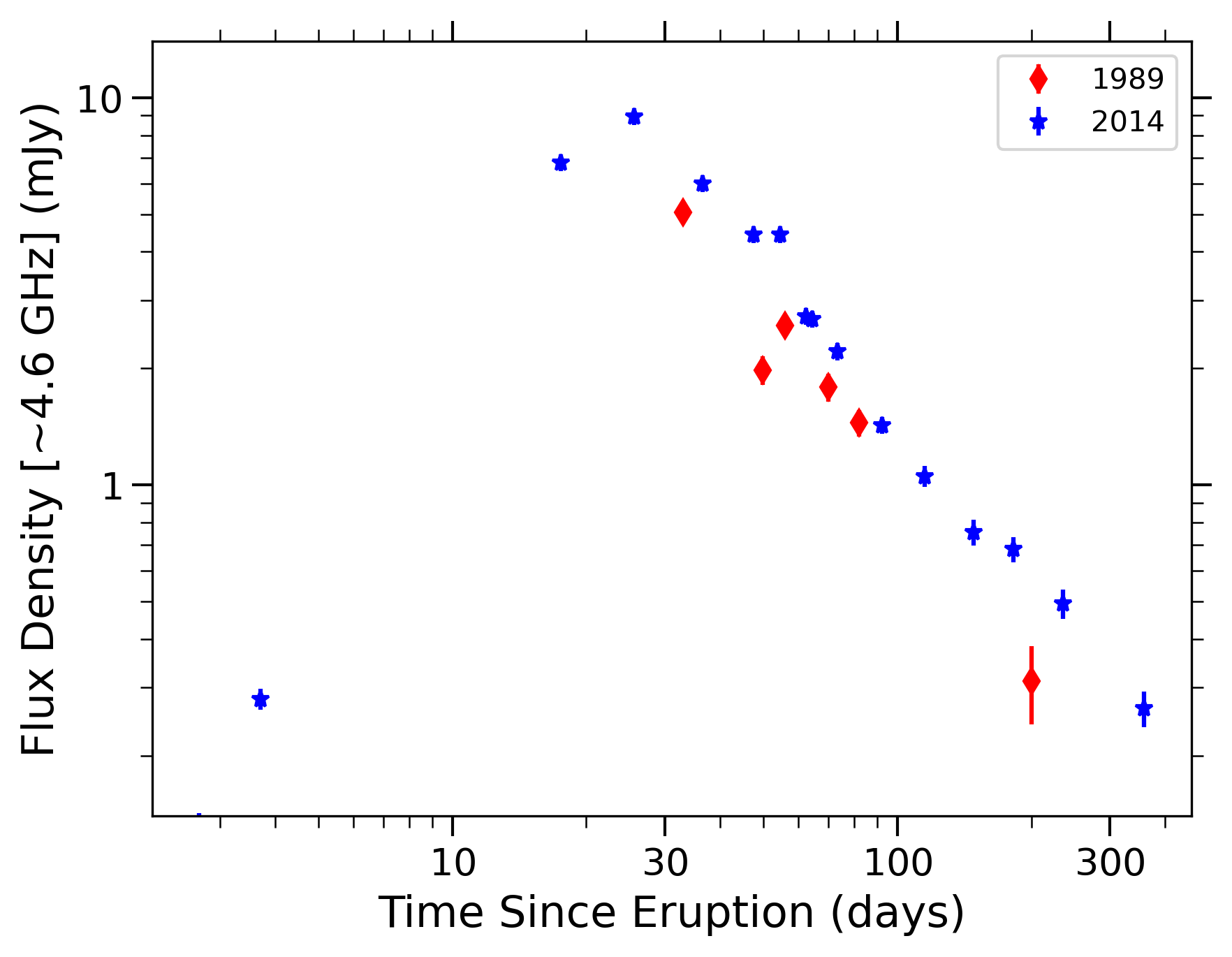}
    \caption{A comparison of the radio light curves of the 1989 and 2014 eruptions of V745 Sco, observed at 4.9 GHz in 1989 (diamonds) and 4.6 GHz in 2014 (stars). Measured flux densities are plotted with error bars,and 3$\sigma$ upper limits as upside down triangles.} 
    \label{fig:c_comp}
\end{figure}

\begin{figure}
    \centering
    \includegraphics[width=85mm]{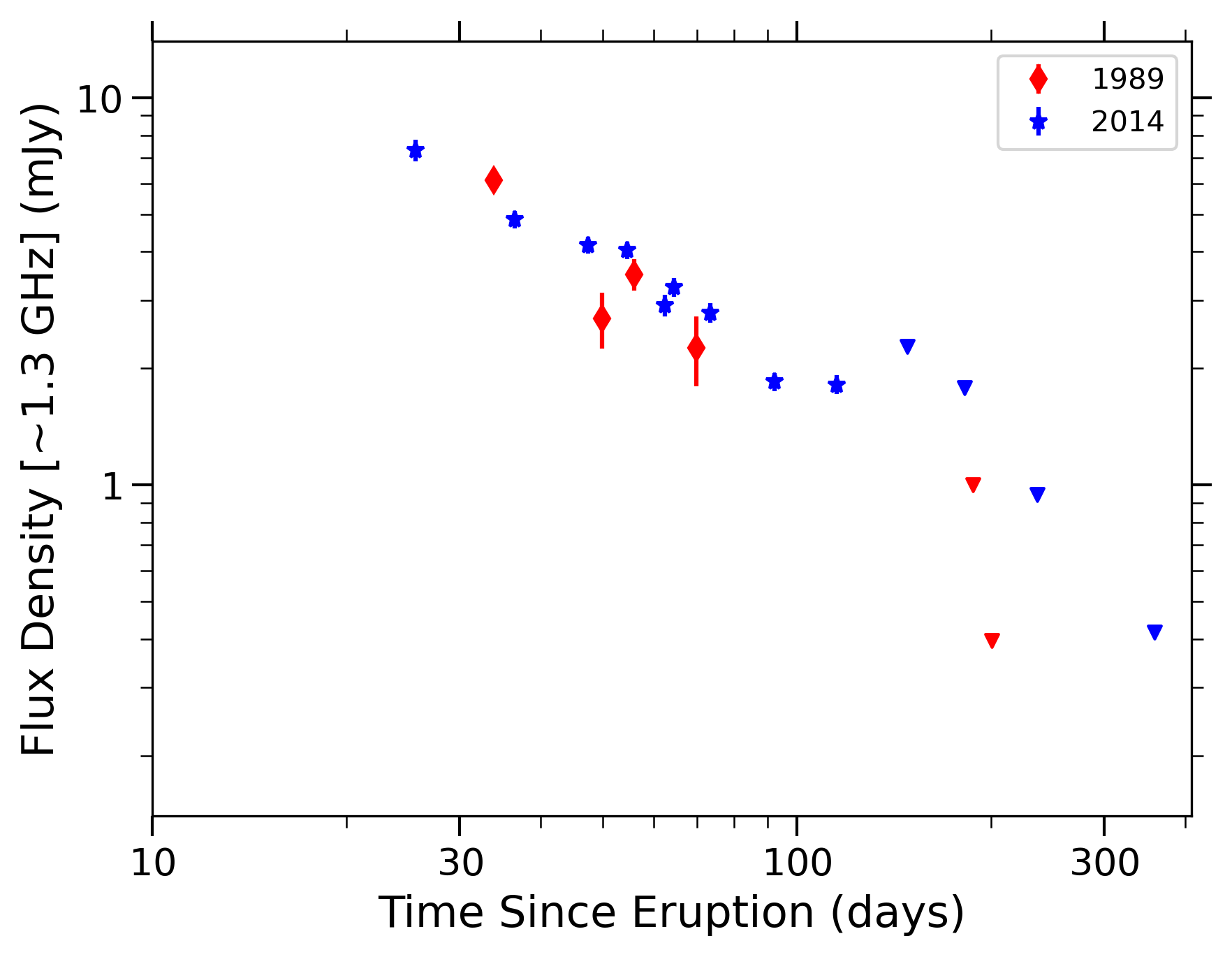}
    \caption{A comparison of the radio light curves of the 1989 and 2014 eruptions of V745 Sco, observed at 1.5 GHz in 1989 (diamonds) and  1.3 GHz in 2014 (stars). Measured flux densities are plotted with error bars, and 3$\sigma$ upper limits as upside down triangles.}
    \label{fig:l_comp}
\end{figure}

The 1989 radio light curve does not exactly follow the same path as the 2014 radio light curve, but they are similar. Figure~\ref{fig:c_comp} compares the radio light curves of the two eruptions as observed $\sim$4.9 GHz, showing that coverage of the 1989 eruption does not start until after light curve peak. We also have fewer observations for the 1989 eruption and larger error bars. At 4.6 GHz, the 2014 light curve declines from peak at a rate of $S_{\nu} \propto t^{-1.21 \pm 0.08}$, while the 1989 light curve declines at a rate of $ t^{-1.47 \pm 0.19}$. For the 4.9 GHz band, we would need to scale the 1989 light curve by 1.2 to get it to align with the 2014 light curve. The discrepancy between the 1989 and 2014 light curves is unlikely to be explained by refining the time of eruption, as the start of eruption is well measured in both cases, with $<$1 day uncertainty \citep{Schaefer10, Waagen14}. Similar behaviour is seen at $\sim$1.5 GHz, with the 1989 data  at slightly lower flux densities than the 2014 data (Figure~\ref{fig:l_comp}). The 2014 1.3 GHz light curve declines at a rate of $S_{\nu} \propto t^{-1.05 \pm 0.21}$, and the 1989 1.5 GHz light curve declines at a rate of  $t^{-1.28 \pm 0.47}$. 

\citet{Schaefer10} has shown that the optical light curves for subsequent eruptions in recurrent novae are similar, but comparable analyses are sparse at radio wavelengths. \citet{Kantharia+16} compared the 1.4 GHz light curve of V745 Sco's 1989 eruption (similar to what is published here) with a 610 MHz light curve of the 2014 eruption (observed with the GMRT; also plotted in Figure~\ref{fig:lcKanth2014}). Fitting the light curves with a parametric model of synchrotron emission developed for radio supernovae \citep{Weiler+02}, they find that the radio emission becomes unabsorbed and peaks earlier during the 2014 eruption, as compared to the 1989 eruption. \citet{Kantharia+16} argue that this reflects evolution in the CSM over decades. Our observations improve on this comparison by observing the two eruptions at similar frequencies, enabling a direct comparison without modelling of the eruption. We do not find any evidence of the 2014 light curve peaking earlier than the 1989 light curve. 
Based on the observations in hand, we can conclude that the period of decline from light curve maximum, $\sim 30-200$ days following eruption, was fainter for the 1989 eruption than for the 2014 eruption. This is the period when the radio optical depth should be relatively low, so it is unlikely that changes in the absorbing medium (due to i.e., a different orientation of the binary between the two eruptions) can explain the difference. 
It is instead likely that the deviation between the radio light curves is reflecting real differences in the energetics of the synchrotron-emitting material between eruptions, either implying a faster shock or denser CSM being shocked in 2014.
Similar conclusions were reached by \citet{Nayana+24}, who compared 0.3--1.3 GHz observations of the 2006 and 2021 eruptions of RS Oph, and found that the 2021 eruption was systematically brighter at radio wavelengths. 

\subsection{Brightness Temperature Evolution} \label{sec:temp}
We estimated the radio brightness temperature ($T_B$) of V745 Sco over the course of its 2014 eruption to test if the radio emission is dominated by thermal or non-thermal emission. A brightness temperature exceeding a value of $5 \times 10^{4}$ K cannot be produced by photo-ionized gas \citep{Cunningham+15},
and the radio emission expected for thermal emission from the hot shocked X-ray-emitting gas will be optically thin and below the detection threshold of the VLA \citep{Weston+16a, Weston+16b, Sokolovsky+23}. 
The implication is that a measurement of $T_B > 5 \times 10^{4}$ K 
indicates non-thermal radio emission \citep{Chomiuk+21radio}.

Calculating brightness temperature requires the flux density at a specific frequency and the angular size of the emitting source:
\begin{equation} \label{eq:BT}
T_B =  1763.1\,K \left(\frac{\nu}{\rm GHz}\right)^{-2} \left(\frac{S_{\nu}}{\rm mJy}\right) \left(\frac{\theta}{\rm arcsec}\right)^{-2} 
\end{equation}
Here, $\theta$ represents the angular diameter of the emitting source, assuming the source is disc-shaped. Ideally, $\theta$ would be directly estimated from imaging, but Very Long Baseline Array (VLBA) data of V745 Sco's 2014 eruption could not be calibrated due in part to instrumental issues and the source's low elevation which limited what antenna could continuously observe the nova. Unfortunately, radio images were not salvageable. Instead, we inferred angular size measurements for the nova ejecta, assuming a distance of 8.2 kpc and a time-dependent physical diameter of the ejecta, calculated assuming ejection on 2014 Feb 6.7 and a constant expansion velocity of $V = 4450$ km s$^{-1}$ (FWZI/2 for the first H$\alpha$ observation on day 3). Note that this is almost certainly an upper limit on the diameter of the ejecta, as they will decelerate upon interaction with CSM, so resulting $T_B$ estimates are lower limits (the same thing is true if the synchrotron emission does not originate over the full ejecta, but instead is limited to compact knots; the resulting $T_B$ if imaged with e.g. VLBA would be significantly higher than that estimated here).

\begin{figure}
    \centering
    \includegraphics[width=80mm]{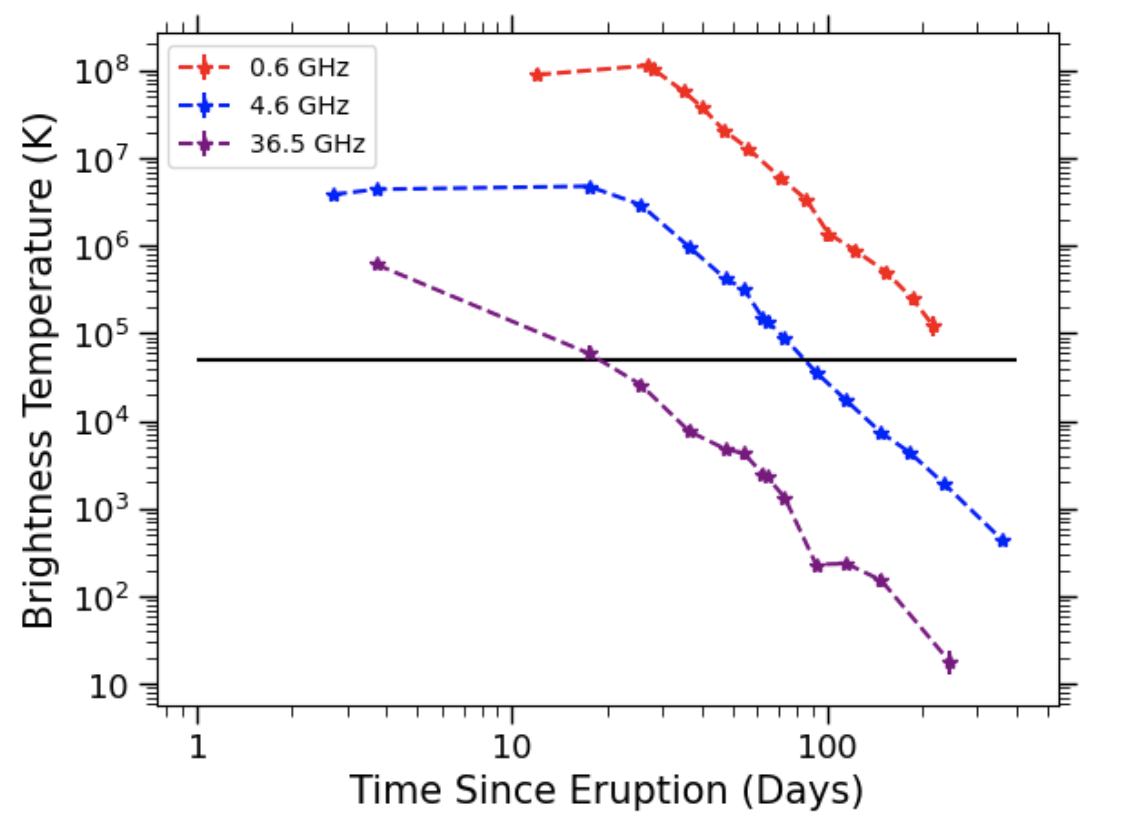}
    \caption{Brightness temperature as a function of time following V745 Sco's 2014 eruption. The brightness temperature is estimated at three frequencies: 0.6 GHz (red) \citep{Kantharia+16}, 4.6 GHz (blue) and 36.5 GHz (purple). Each point on the dashed lines correspond to an observation obtained in the corresponding frequency band. The horizontal black line indicates where $T_B=5 \times 10^{4}$K. }
    \label{fig:tb}
\end{figure}

Figure~\ref{fig:tb} plots the calculated brightness temperature as a function of time covering the 2014 eruption at three different frequencies. At 0.6 GHz \citep{Kantharia+16}, the brightness temperature starts out around $10^8$ K, and remains above  $5 \times 10^{4}$ K over the course of the eruption, for more than 200 days. This indicates that the emission is synchrotron dominated at all times at the lower frequencies. 
This is consistent with the analysis of \citet{Kantharia+16}, who assume the lower frequency radio emission ($0.2 - 1.4$ GHz) from V745 Sco is synchrotron radiation. 
Constraints on $T_B$ from higher frequency observations are not as stringent as the 0.6 GHz data (Figure~\ref{fig:tb}); the $T_B$ estimate drops below $5 \times 10^{4}$ K by day 92 at 4.6 GHz, and by day 25 at 36 GHz.  Therefore, it is possible that thermal emission contributes to higher frequencies at later times. 
However, note that the brightness temperature expected for optically-thin thermal gas is significantly lower than the physical temperature of the gas ($T$), such that $T_B \approx \tau T$, where $\tau$ is the optical depth in the optically thin regime. The radio spectrum is optically thin at these late times (\S \ref{sec:spec}), so it is doubtful that thermal emission can be the dominant radio emission mechanism at e.g. 4.6 GHz on day 200. 

\subsection{Radio Spectral Evolution} \label{sec:spec}

\begin{figure}
    \centering
    \includegraphics[width=80mm]{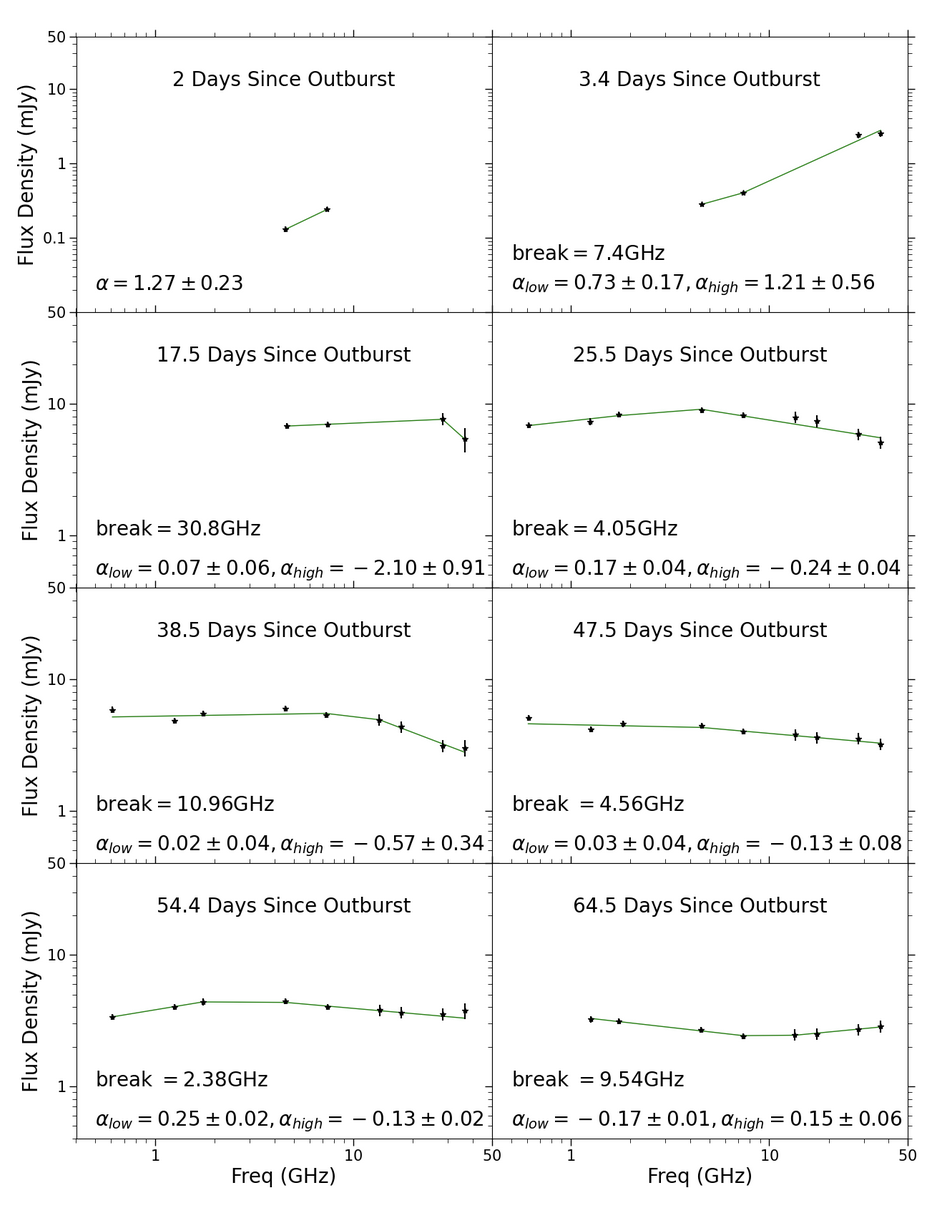}
    \caption{Radio spectra observed over 8 early epochs of V745 Sco's 2014 eruption. Each row has a fixed y-axis range. Power-law fits ($S_{\nu} \propto \nu^{\alpha}$) to the spectra are overplotted, and the resulting spectral indices are listed in each panel. Data from \citet{Kantharia+16} was included.}
    \label{fig:spec_fit1}
\end{figure}

\begin{figure}
    \centering
    \includegraphics[width=80mm]{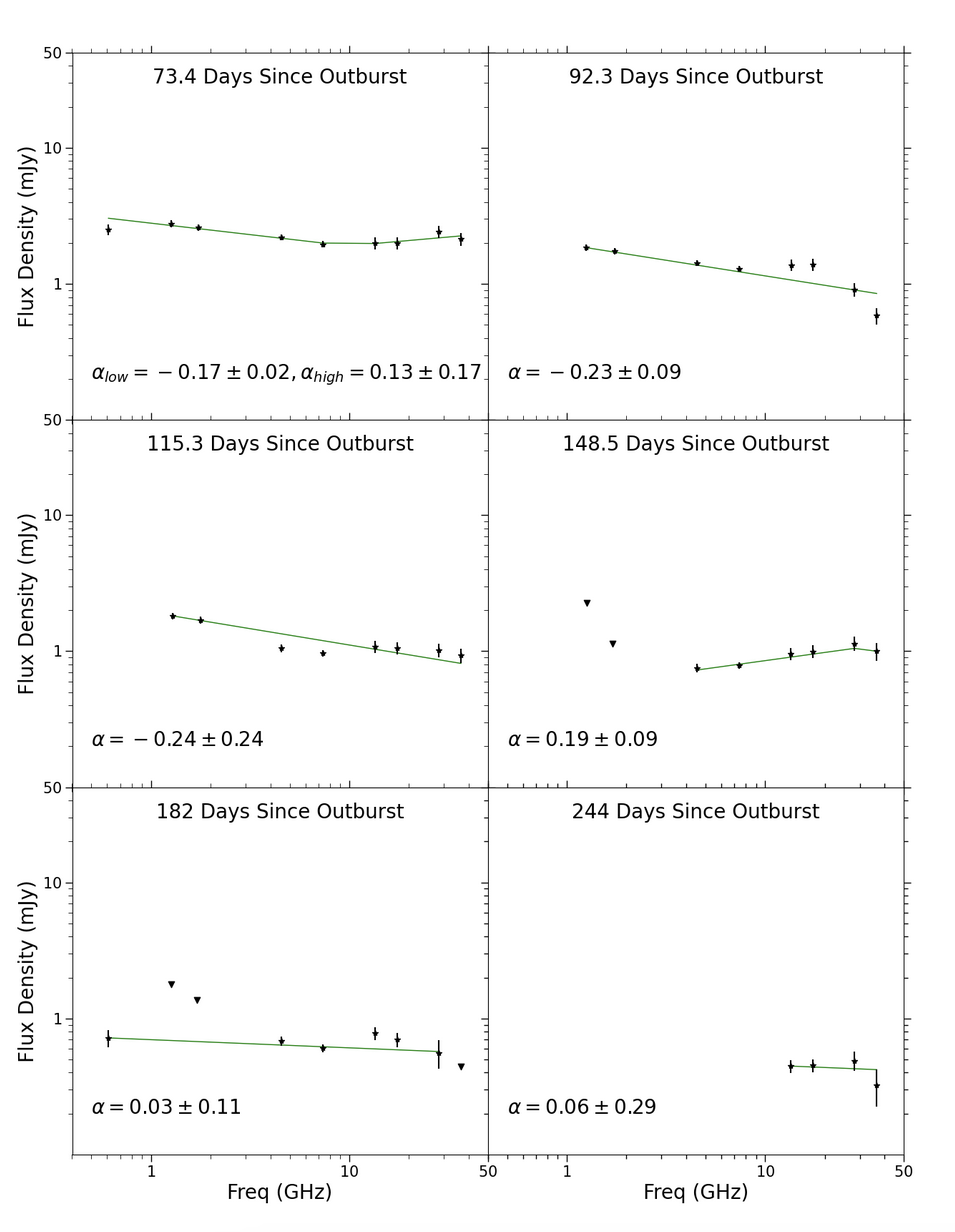}
    \caption{Radio spectra observed over 6 later epochs of V745 Sco's 2014 eruption. 3$\sigma$ upper limits are plotted as black triangles. Each row has a fixed y-axis range. Power-law fits to the spectra are overplotted, and the resulting spectral indices are listed in each panel.}
    \label{fig:spec_fit2}
\end{figure}

To assess how the radio emission evolves over time and how long it remains optically thick, we analysed the radio spectral evolution of V745 Sco over the course of its eruptions. Figure~\ref{fig:spec_fit1} and Figure~\ref{fig:spec_fit2} plot flux density against frequency for each radio epoch during the 2014 eruption. A best fit line (in log--log space) is plotted over the spectra, the slope of which yields the spectral index. 
The spectral index $\alpha$ is a measure of how the flux density depends on frequency, given by the relation
$S_{\nu} \propto \nu^{\alpha}$.
For most epochs, a broken power law was used to accommodate different slopes at low and high frequencies, while some observations were better fit with a single power law. The reduced $\chi^2$ value of the fits was used to decide if a single or broken power law fit the data better. The broken power law requires a break point, the frequency at which one power law ends and the other begins. A range of break points were tried, and the best fit determined by minimizing the reduced $\chi^2$ value. 
$\alpha_{high}$ corresponds to the slope of the frequencies above this break point, and $\alpha_{low}$ corresponds to the slope of the frequencies below the break point. These fits to the spectra were made using  LevMarLSQfitter in Astropy, a fitter which uses a Levenberg-Marquardt algorithm to optimize the fit. Flux measurements input into the power law fitter were weighted by $\frac{1}{error^2}$. Upper limits were not included in the broken power law or the fit.

\begin{figure}
    \centering
    \includegraphics[width=80mm]{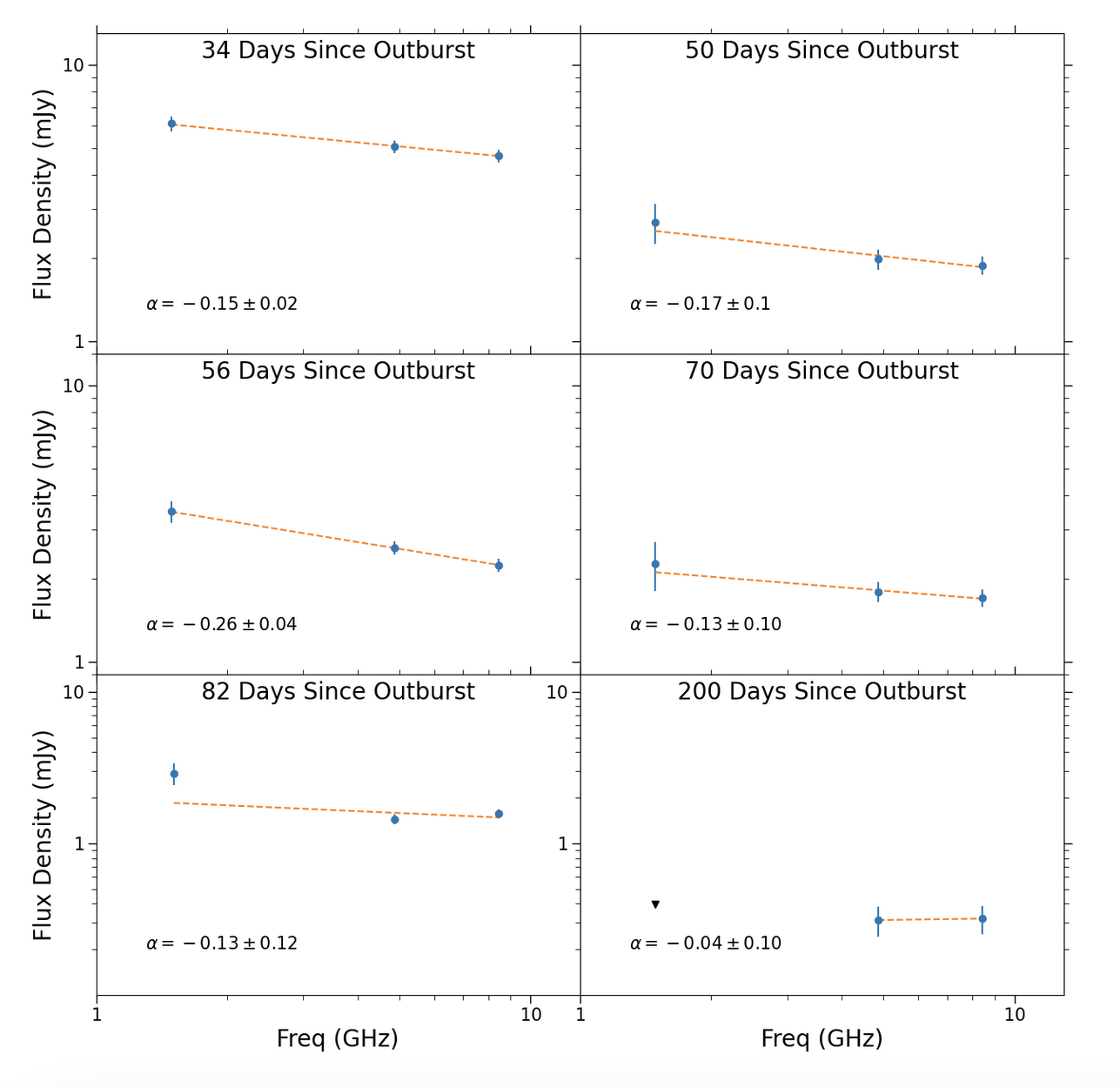}
    \caption{Radio spectral evolution of the 1989 eruption. 3$\sigma$ upper limits are plotted as black triangles. Each row has a fixed y-axis range. Power-law fits to the spectra are overplotted, and the resulting spectral indices are listed in each panel.}
   
    \label{fig:1980S}
\end{figure}

Before radio peak, the ejecta are optically thick and the spectrum rises steeply ($\alpha \approx 1.3$ on day 2). Around radio peak on day 25, the spectrum is relatively flat below $\sim$10 GHz ($ \alpha = 0.2$), and decreases with frequency at higher frequencies ($ \alpha \approx -0.2$). This behaviour is consistent with radio emission transitioning to an optically thin state, where $\tau \approx 1$ is around 10 GHz. Similar spectral shapes continue for the next three observations, through day 54. Generally, we expect higher frequencies to become optically thin earlier, because it takes more material to be optically thick at high frequencies. 
Around day 64, the spectrum changes shape from a concave to a convex morphology, with a mildly negative spectral index ($\alpha = -0.2$) at lower frequencies, and a mildly positive spectral index ($\alpha = 0.2$) at higher frequencies. It is possible there is a combination of thermal emission at higher frequencies and non-thermal emission at lower frequencies. This behaviour continues for day 72. For the last five observations, the spectrum is flat for all frequencies ($\alpha \approx 0.0$). 

Figure~\ref{fig:1980S} shows the spectral evolution of V745 Sco's 1989 eruption. The first epoch of the 1989 eruption, 34 days into eruption, shows the lower frequencies peaking at slightly higher flux density values than the higher frequencies, with $\alpha = -0.15 \pm 0.02$. The slope is most steep on day 56 ($\alpha = -0.26$) and then remains relatively flat ($\alpha = -0.13$). There are fewer points in the 1989 spectral index plots so we cannot tell how the spectral index evolves at higher frequencies ($>$ 8.44 GHz). The slope of the lower frequencies remains negative at all times, and like the 2014 spectra, it is flatter at later times. However, the slope between the 1.49 GHz and 4.86 GHz measurements is never positive, as we see for the 2014 eruption before day 64. 

\begin{figure}
    \centering
    \includegraphics[width=80mm]{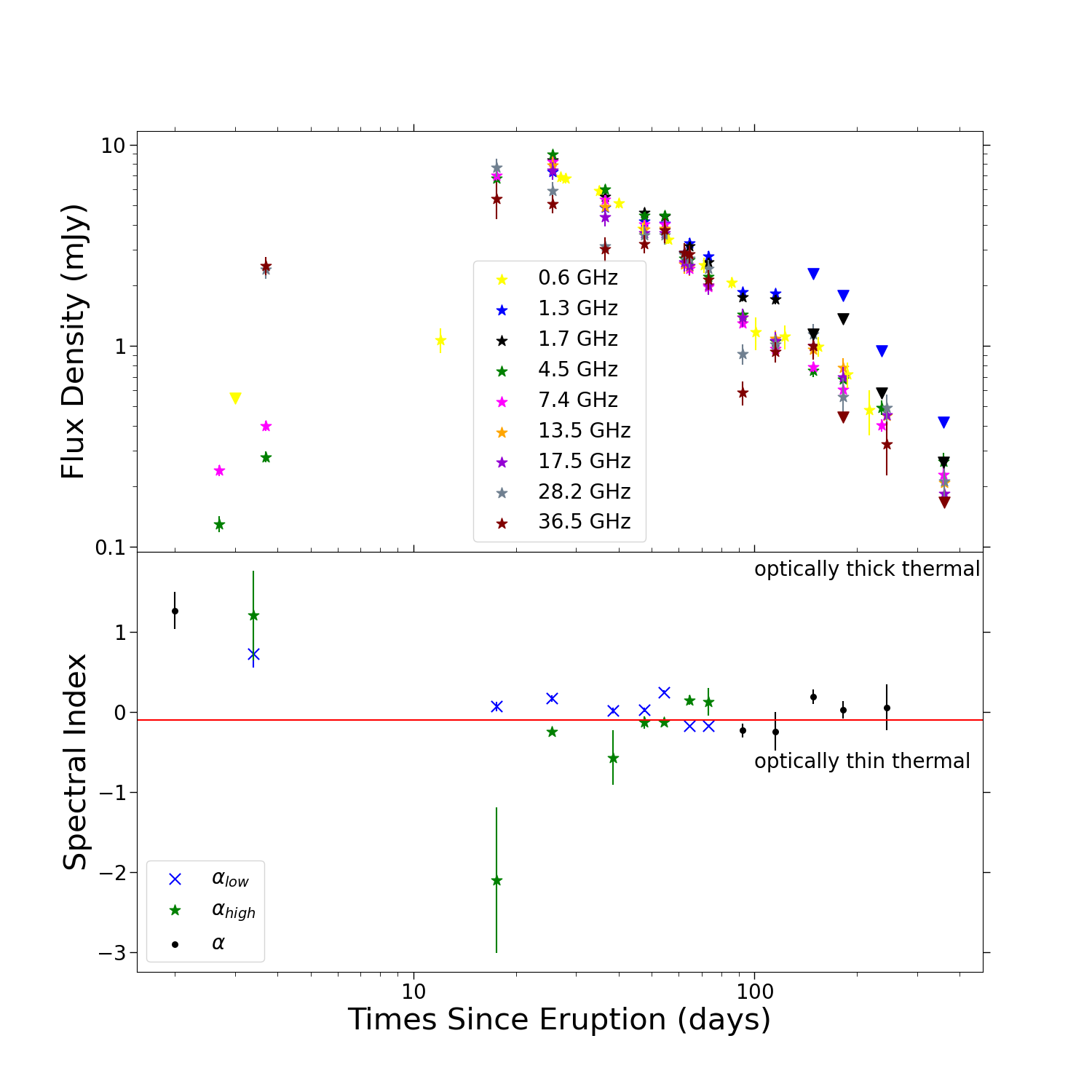}
    \caption{Stacked plots of the multi-frequency radio light curve (top panel) and the spectral index as a function of time (bottom panel) for V745 Sco's 2014 eruption. Spectral indices fit to the higher frequencies ($\alpha_{high}$) are plotted as green stars, fits to lower frequencies ($\alpha_{low}$) are plotted as blue x marks, and epochs where one power law can be fit to the whole frequency have spectral indices ($\alpha$) plotted as black points. The alpha value associated with optically thin thermal emission is plotted as a red horizontal line at -0.1.}
    \label{fig:spind}
\end{figure}

The radio spectral evolution of V745 Sco's 2014 eruption is summarized in Figure~\ref{fig:spind}, which shows how the spectral index changes over time for high and low frequencies. It is plotted under the multi-frequency radio light curve to show how the spectral index relates to the radio peak and decline. It shows that the spectral index begins at a value of $\alpha = 1.3$, indicative of optically thick emission. The spectral index then drops quickly and stays around 0, indicative of optically thin thermal emission or partially optically thin synchrotron emission (e.g. \citealt{Blandford&Konigl79}). 

We can compare V745~Sco's radio spectral evolution with the spectral evolution of other novae with synchrotron-dominated light curves. Before radio peak, the spectrum of the symbiotic recurrent nova V3890 Sgr rises with frequency as $\alpha = 1.3$, much as observed in V745~Sco \citep{Nyamai+23}. After radio peak the spectrum turns over and converges to $\alpha = -0.3$, which \citet{Nyamai+23} interpret as indicative of optically thin synchrotron emission. \citet{Linford+17} analysed the spectral evolution of another nova with a giant donor, V1535~Sco. The radio spectrum starts with the flux density decreasing with frequency, as $\alpha \approx -0.9$.
After one of the radio peaks (day 23.7), the spectrum switches to $\alpha \approx 1$. It flattens by day 41 ($\alpha \approx 0.2$), and further flattens at later times (day 78.5, $\alpha = -0.06$). 

The relatively steeply inverted spectra observed at early times ($\alpha \approx 1$ on days 2 and 3 in V745 Sco) require a radio absorption mechanism. Free-free absorption from warm ionized CSM between the observer and the synchrotron-emitting shock is likely, and is expected to decline as the shock moves outward, sweeping up and shock-heating the CSM. 
As discussed in \S\ref{sec:model}, synchrotron self-absorption is unlikely to be important in V745 Sco. Alternatively, \cite{Taylor+87} suggested the Razin-Tsytovich effect
suppressing synchrotron radiation in the presence of thermal plasma 
\citep{1965ARA&A...3..297G,Rybicki&Lightman79} may be responsible for a high spectral index of a nova. 
This effect has also been discussed in the context of  colliding-wind binaries \citep{2006A&A...446.1001P}, 
synchrotron emission from winds of single hot stars \citep{2004A&A...418..717V,2022ApJ...932...12E}
and active galactic nuclei \citep{1966AuJPh..19..195K,1998ApJ...504..147R}. For any absorption mechanism, a gradient in the optical depth across the source may flatten the spectrum compared to a uniform source affected by this mechanism.

After light curve peak, novae regularly show shallower spectral indices for their synchrotron emission than the $\alpha = -0.5$ to $-1$ typically expected for diffusive shock acceleration \citep{Bell78, Blandford&Ostriker78} and observed for optically thin synchrotron emission in other astrophysical phenomena \citep[e.g.][]{Weiler+02,Green19}---and V745 Sco is no exception. However, brightness temperature arguments (\S \ref{sec:temp}) confirm that the emission is in fact non-thermal, at least at the lower frequencies (Figure~\ref{fig:tb}). 
The radio emission we observe in novae is likely a combination of thermal and non-thermal emission. The thermal and synchrotron components may change the slope of the observed spectrum or produce a more complex spectral shape.
However, even imaging with Very Long Baseline Interferometry (VLBI), 
which is sensitive only to high brightness temperature synchrotron emission, 
reveals relatively flat spectral indices in these novae \citep{Rupen+08, Chomiuk+14,
Linford+17}, suggesting that the radio spectrum of synchrotron emission in novae may be unusually flat. If thermal emission contributes to the radio light curve of V745 Sco, 
it is likely at relatively high frequencies and at relatively late times, perhaps dictating `$\alpha_{\rm high}$' after day 60 
(the estimated brightness temperature at 36.5 GHz on day 64 is $\sim 10^{4}$, which cannot exclude thermal emission).

Very likely, complex optical depth effects are also playing a part in producing the flat radio spectra of non-thermal emission in novae \citep[e.g.][]{Vlasov+16}. Flat spectral indices are often observed in X-ray binaries in quiescence and many active galactic nuclei \citep[e.g.][]{Fender01, Nagar+02}, 
and are usually interpreted as partially optically thin synchrotron emission from a self-absorbed compact conical jet
\citep{Blandford&Konigl79,1981ApJ...243..700K}. 
The key difference between a quiescent X-ray binary jet and a nova eruption is that the former can drive a steady outflow of plasma, yielding a self-similar geometry that can maintain a flat radio spectrum over significant frequency and time ranges, while novae confine mass ejection to a relatively short period of time during the eruption. Because the ejecta in V745 Sco are expanding and the density of the synchrotron-emitting regions (and any absorbing screen) is greatly dropping over the time period covered by our radio observations, one would therefore expect the radio spectrum to converge to an optically thin 
value at late times. Indeed, when X-ray binaries enter an outbursting state, they are observed to launch synchrotron-emitting blobs with steep radio spectra, as expected for optically thin synchrotron emission \citep{Fender01}---but no such transition to a steeper radio spectrum is seen in V745 Sco (or other novae). 

There is also the possibility that the energy spectrum of relativistic electrons in novae diverges from 
the $N(E) \propto E^{-2}$ power law typically expected from diffusive shock acceleration \citep{Bell78, Blandford&Ostriker78}, and is instead significantly flatter. 
Flatter energy spectra for relativistic particles are  
predicted by non-linear diffusive shock acceleration models, wherein a significant fraction of the post-shock energy density is transferred to relativistic particles, leading to higher shock compression ratios \citep[e.g.][]{Caprioli23}.

Another possibility is that there is a low-energy cutoff in the relativistic electron spectrum above the frequencies where the slightly positive $\alpha$ is measured. 
A spectral index of $\alpha = 0.3$ is expected below the low-energy cutoff frequency, 
which is the slope of the synchrotron spectrum of a single electron
\citep{1965ARA&A...3..297G,Rybicki&Lightman79}.
This might explain the concave but relatively flat spectra during days 25--54 and the measured values of $\alpha_{\rm low}$ during this time. 
In this scenario, the low-energy cutoff moves to lower energies as the eruption evolves, dropping below our observed frequency by day 64.

\section{Modelling the Radio Light Curve} 
\label{sec:radiolcmodel}

If we have an estimate of the blast wave dynamics and CSM density profile that the blast is expanding into, it is possible to use the formalism pioneered by \citet{Chevalier82} to predict the synchrotron emission. Our goal is to compare the synchrotron model with our radio observations of V745 Sco, in order to constrain the properties of the CSM surrounding V745 Sco (\S \ref{sec:model}). In \S \ref{sec:snia}, we compare V745 Sco's CSM with constraints on CSM around SNe Ia (primarily also from radio observations), in order to test V745 Sco as a viable progenitor system for SNe Ia.

\subsection{A Model for Synchrotron Emission from V745 Sco} \label{sec:model}

Here, we consider the synchrotron emission that might be expected from a simple nova ejecta interacting with a spherically symmetric wind-like CSM described by Equation \ref{eq:winddens}.  
We set $v_{w}$, the wind velocity of the red giant, to 10 km s$^{-1}$ throughout this paper. 
The ejecta are assumed to expand homologously ($v_{ej} \propto r$) and to have a density profile $\rho_{ej} \propto r^{-2}$ \citep{Hjellming+79, Hauschildt+92}. The dynamics of this interaction are described by \citet{Tang&Chevalier17} over the course of the free expansion and Sedov-Taylor phases. As described in \S \ref{sec:linewid}, we take an ejecta kinetic energy of $5 \times 10^{42}$ erg ($5 \times 10^{43}$ erg) for a $10^{-7} M_{\odot}$ ($10^{-6} M_{\odot}$) ejection, to match the early shock velocity to the early H$\alpha$ FWZI/2.

The model for synchrotron emission is very similar to that applied to the recurrent nova V3890 Sgr by \citet{Nyamai+23}. In the shock between the nova ejecta and CSM, particles are accelerated to relativistic speeds due to the Fermi mechanism \citep{Blandford&Ostriker78}, and the magnetic field is amplified due to the streaming instability \citep{Bell04}. The energy densities in the post-shock magnetic field ($U_B$) and in relativistic electrons ($U_e$) are assumed to be fractions of the post-shock energy density:
\begin{equation} \label{eq:ub}
U_B = \epsilon_B\,  \rho_{CSM}\,  v^2_s
\end{equation}
\begin{equation} \label{eq:ue}
U_e = \epsilon_e\,  \rho_{CSM}\,  v^2_s
\end{equation}
Here, $\epsilon_B$ is the fraction of the post shock energy density in the magnetic field and  $\epsilon_e$ is the fraction of the shock power in relativistic leptons. 
The microphysical parameters $\epsilon_e$ and $\epsilon_B$ are uncertain, and may depend on the Mach number of the shock, the inclination angle, and strength of the magnetic field \citep{Vlasov+16}. 

The energy spectrum of relativistic electrons is assumed to be a power law:  $N(E) \propto E^{-p}$. For optically thin synchrotron emission, the radio spectral index is related to the electron spectrum as $\alpha = (1-p)/2$. Taking an $\alpha=-0.2$ (measured when the radio emission is optically thin in V745 Sco) gives  $p=1.4$. However, the formalism of \citet{Chevalier98} is only valid for $p>2$, and it is not clear at this time if the spectra of relativistic electrons in novae are in fact unusually flat (see discussion in \S \ref{sec:spec}). 
Therefore,  we take $p=2.1$ in our analysis. The corresponding constants for calculating the synchrotron luminosity are then $C_5 = 1.37 \times 10^{-23}$ and $C_6 = 8.61 \times 10^{-41}$ \citep{Pacholczyk70}. 

As described by \citet{Nyamai+23} (their Equation 3), we use the equations of \citet{Chevalier98} to predict the radio synchrotron flux density at an arbitrary frequency $\nu$, accounting for synchrotron self-absorption. 
From Equation 1 of \citet{Chevalier98} and Equations \ref{eq:ub} and \ref{eq:ue} above, the optical depth to synchrotron self absorption is
\begin{equation}
\tau_{\rm SSA} \propto \sqrt{
{\epsilon_e}^{2}\, \epsilon_B^{(p+2)/2}\, \rho_{\rm CSM}^{(\frac{1}{2}p+3)}\, v_s^{(p+6)}\over{\nu^{(p+4)}}}
\end{equation}
In the case where $p=2$, the proportionality simplifies to $\tau_{\rm SSA} \propto \epsilon_e \epsilon_B \rho_{\rm CSM}^2 v_s^4 \nu^{-3}$. This demonstrates that faster shocks and shocks with denser CSM will suffer stronger synchrotron self-absorption. The relatively slow shocks of novae ($\sim$3000 km s$^{-1}$) compared to e.g. Type Ibc SNe ($\sim$30,000 km s$^{-1}$) mean that synchrotron self-absorption is usually negligible in novae (while it can be important in Type Ibc SNe; \citealt{Chevalier98,Chevalier&Fransson06}), and this is what our models predict for parameters consistent with V745 Sco. 

A much more important source of radio opacity in V745 Sco is free-free absorption from the warm ionized CSM between the shock and the observer. We approximate $\tau_{ff}$ by again assuming a spherical wind-like CSM, maintained at a photoionized temperature of $10^4$ K, and using Equation 14 from  
\citet{Chevalier81}:
\begin{align}
\begin{split}
\tau_{\textrm{ff}} &= 0.005
\bigg(\frac{{\dot M}}{10^{-6}~\textrm M_{\odot}~\textrm{yr}^{-1}}\bigg)^{2} \times
\bigg(\frac{v_\textrm {w}}{10~\textrm{km~s}^{-1}}\bigg)^{-2} \times \\
&\bigg(\frac{\nu}{1.4~\textrm {GHz}}\bigg)^{-2} \times \bigg(\frac{R}{2 \times 10^{16}~\textrm{cm}}\bigg)^{-3}
\label{eq:tauff}
\end{split}
\end{align}
where $R$ is the radius of the shock. The light curves at frequencies, 0.6--17.5 GHz, all peak around day 25 (Figure~\ref{fig:lcKanth2014}), implying that $\tau_{ff} \approx 1$ at this time (at all these frequencies near simultaneously). Running \citet{Tang&Chevalier17} models for reasonable parameters of the inner CSM (constrained by X-ray observations of \citealt{Delgado&Hernanz19}: $\dot{M} = [5-10] \times 10^{-7}\ \textrm M_{\odot}~\textrm{yr}^{-1}$, $v_{w} = 10$ km s$^{-1}$), in combination with $M_{ej} = 10^{-7}-10^{-6}$ M$_{\odot}$, and kinetic energies as derived in \S\ref{sec:linewid}, the shock radius on day 25 is $[1.3-3.2] \times 10^{14}$ cm. Meanwhile, if the CSM is asymmetric (e.g., largely confined in an equatorial density enhancement), the shock may be relatively undecelerated in some directions, and could reach $9.6 \times 10^{14}$ cm by day 25 (expanding at 4450 km s$^{-1}$.
From Equation \ref{eq:tauff} and taking the most constraining frequency of 0.6 GHz, we can infer that the CSM external to this radius is characterized by $\dot{M} < [3-64] \times 10^{-9}\ \textrm M_{\odot}~\textrm{yr}^{-1}$ (assuming $v_{w} = 10$ km s$^{-1}$). This $\dot{M}$ is one to two orders of magnitude less than the $[5-10] \times 10^{-7}\ \textrm M_{\odot}~\textrm{yr}^{-1}$ implied by the X-ray absorbing column in the eruption's first few days \citep{Delgado&Hernanz19}, in line with the suggestion by Delgado \& Hernanz that the inner dense CSM is truncated beyond several $\times 10^{14}$ cm. The radio and X-ray data can be reconciled if the CSM within this truncation radius is characterized by $\dot{M}_{in} = [5-10] \times 10^{-7}\ \textrm M_{\odot}~\textrm{yr}^{-1}$, and the outer CSM is significantly less dense: 
$\dot{M}_{out} \la 6 \times 10^{-8}\ \textrm M_{\odot}~\textrm{yr}^{-1}$. 
In this case, $\dot{M}_{out}$ would produce a small amount of absorption in the X-ray, $N_H \la 1 \times 10^{20}$ cm$^{-2}$, compared to the larger ISM column, $\sim 6 \times 10^{21}$ cm$^{-2}$ (\S \ref{sec:dist}; see also \citealt{Page+15} and \citealt{Delgado&Hernanz19}).

\begin{figure}
    \centering
    \includegraphics[width=80mm]{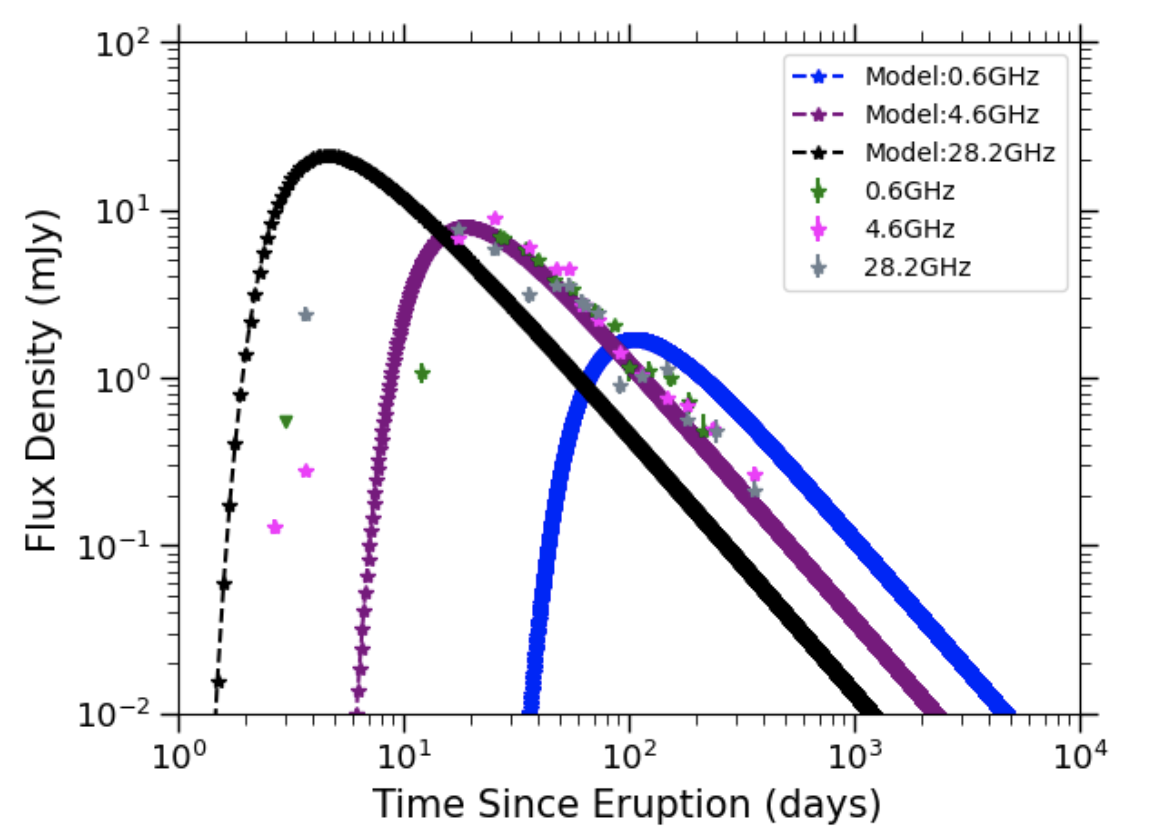}
    \caption{Modelled synchrotron emission light curves at three frequencies: 0.6 GHz (blue), 4.6 GHz (purple) and 28.2 GHz (black), superimposed on the observed radio light curve at corresponding frequencies. The model was selected to fit the 4.6 GHz light curve, and takes $\dot{M} = 10^{-7}$ M$_{\odot}$ yr$^{-1}$, $M_{ejecta} = 10^{-7}$ M$_{\odot}$, $\epsilon_e =\epsilon_B = 0.0028$. At different frequencies, this model peaks on different days and at different flux densities, in a matter that is inconsistent with observations.}
    \label{fig:synch}
\end{figure}

The need for a truncated inner dense CSM can be seen in  Figure~\ref{fig:synch}, which shows the modeled synchrotron emission on top of the observed light curves of V745 Sco's 2014 eruption for three representative frequency bands. In this model, we fit the observed light curve peak at 4.6 GHz, which yields $\dot{M} = 10^{-7}\ \textrm M_{\odot}~\textrm{yr}^{-1}$; meanwhile, relatively low values of $\epsilon_e = \epsilon_B = 0.0028$ are needed to fit the optically thin decline of the light curve with this relatively high $\dot{M}$.
At a given frequency, at early times, the model light curve rises steeply as the free-free optical depth decreases (due to the blast moving out into the CSM), and peaks when $\tau_{ff} \approx 1$. The light curve then declines as the shock decelerates and interacts with less dense CSM. As expected from Equation \ref{eq:tauff}, we see that the model light curve peaks at 0.6, 4.6, and 28.2 GHz are all distinct from one another, with e.g., 4.6 GHz peaking around day 18, while 0.6 GHz does not peak until after day 100. The observations, on the other hand, show no such behaviour, with 0.6 GHz and 4.6 GHz peaking on day 25, and 28.2 GHz peaking just a bit earlier, on day 17.5. 

The simplest way to obtain a light curve where the different frequencies all peak at the same time is if the synchrotron-emitting shock suddenly `breaks out' from a large absorbing column, so that all frequencies become optically thin in rapid succession. A possible explanation for this discontinuity in the CSM, with a relatively high $\dot{M}_{in}$ within several $\times 10^{14}$ cm and a lower $\dot{M}_{out}$ external, is that the wind from the RG companion has only had time to extend to a certain radius. \citet{Delgado&Hernanz19} suggest that V745 Sco's 1989 eruption swept the environs clean, and the giant has only had ($2014-1989$) = 25 years to pollute the surroundings with its wind. In this scenario, we expect the red giant wind to extend out to $7.8 \times 10^{14}\,(v_{w}/10~{\rm km}\,{\rm s}^{-1})$ cm, which is comparable to within an order of magnitude to the CSM truncation radius estimated here.

Another scenario which could explain the sudden unveiling of synchrotron emission is an asymmetric CSM largely concentrated in the orbital plane \citet{Orlando+17}; if the synchrotron emission is primarily originating in the fastest ejecta expanding in the polar directions, the shock may suddenly appear unabsorbed as it expands beyond the equatorial density enhancement.  To test this idea, in the future we plan to adapt and expand the 3D hydrodynamic model of \citet{Orlando+17} to yield a simulated multi-frequency radio light curve that can be compared with observations. 

\begin{figure}
    \centering
    \includegraphics[width=80mm]{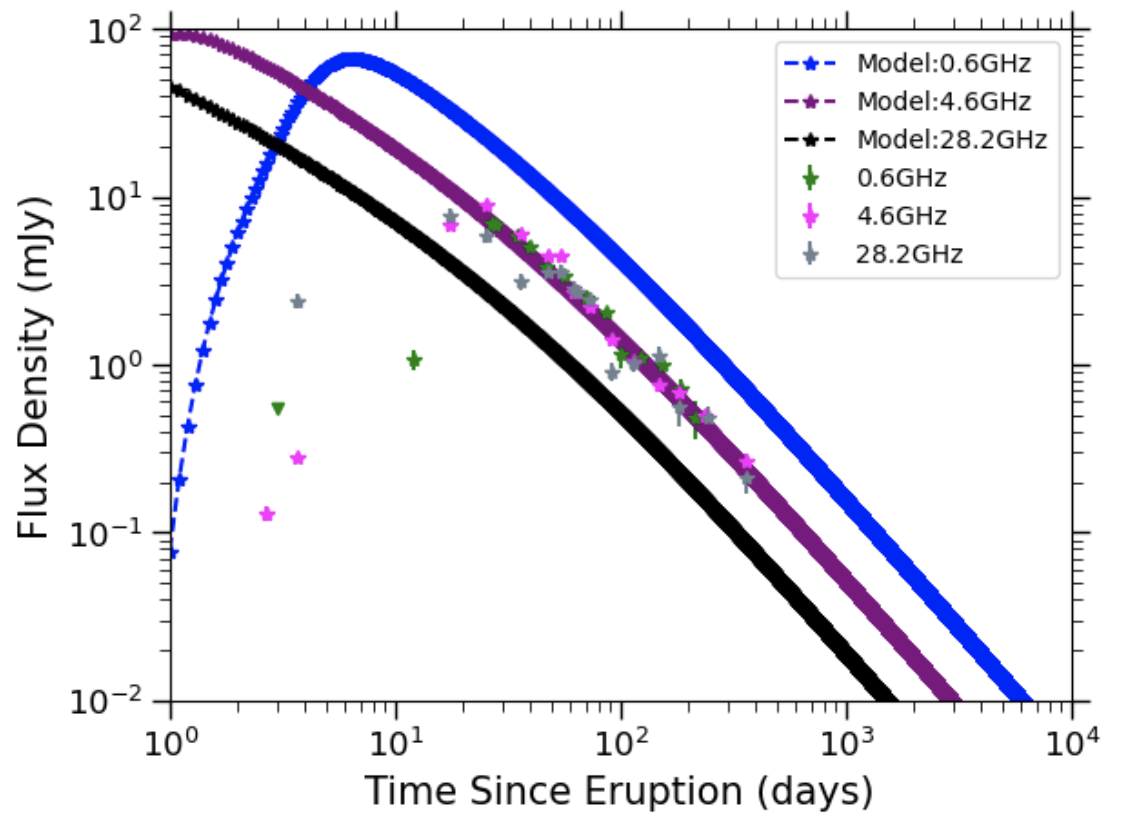}
    \caption{As in Figure~\ref{fig:synch}, synchrotron light curve models are superimposed on observations at three representative frequencies. This model takes $E=5 \times 10^{42}$ erg, $M_{ej} = 10^{-7}$ M$_{\odot}$, $\dot{M} = 9 \times 10^{-10}$ M$_{\odot}$ yr$^{-1}$, $v_w = 10$ km s$^{-1}$, $\epsilon_B = 0.1$ and $\epsilon_e = 0.1$. This model demonstrates that the decline from radio peak can be reasonably well fit by a low $\dot{M}$ and efficient microphysical parameters, if the synchrotron emission until day $\sim$17--28 is absorbed by additional dense CSM at small radii.}
    \label{fig:mod1}
\end{figure}

\begin{figure}
    \centering
    \includegraphics[width=80mm]{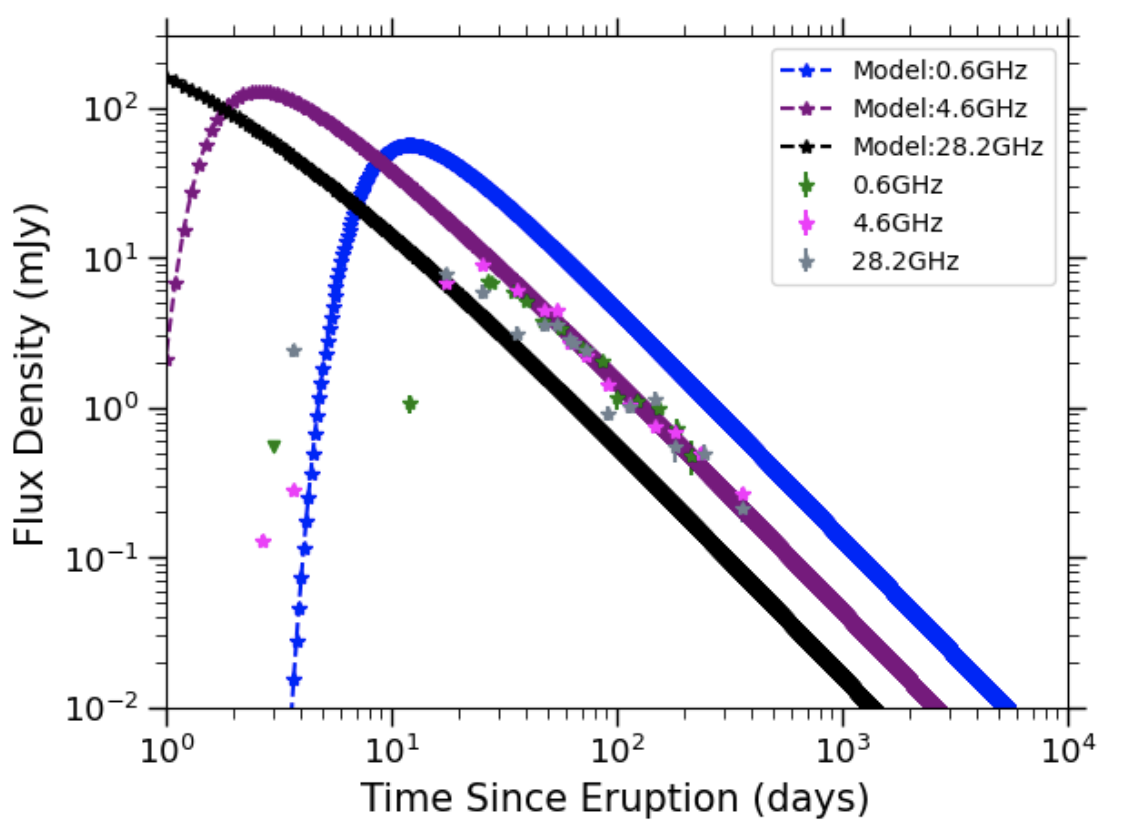}
    \caption{As in Figures~\ref{fig:synch} and \ref{fig:mod1}, synchrotron light curve models are superimposed on three representative frequencies. This model is identical to that in Figure \ref{fig:mod1}, except it takes $\epsilon_B = 0.01$ and $\dot{M} = 7\times 10^{-9}$ M$_{\odot}$ yr$^{-1}$.}
    \label{fig:mod2}
\end{figure}

Beyond day $\sim$25, interaction is ongoing, as is clear from the prolonged tail of synchrotron emission gradually fading out to at least day $\sim$200 (Figure~\ref{fig:lcKanth2014}). This optically thin tail of the light curve can be explained by interaction with a lower density wind-like CSM ($\dot{M}_{out}$). As our goal is to compare V745 Sco's CSM with constraints on CSM around SNe Ia (\S \ref{sec:snia}), we use $\epsilon_e = 0.1$ and $\epsilon_B = 0.1$ and 0.01 to estimate $\dot{M}_{out}$, as \citet{Chomiuk+16} did in their study of SNe Ia at radio wavelengths. Our comparison therefore assumes that SN shocks and nova shocks accelerate particles and amplify $B$ fields with similar efficiency; if $\epsilon_e$ and/or $\epsilon_B$ depend on shock velocity/Mach number (e.g. \citealt{Caprioli&Spitkovsky14}), then it would be more appropriate to use lower $\epsilon$ values in novae; we provide how CSM density scales with $\epsilon_e$ and $\epsilon_B$ at the end of this sub-section. This outer wind can be described by $\dot{M}_{out} =9\times 10^{-10}$ M$_{\odot}$ yr$^{-1}$ for $\epsilon_B = 0.1$ and $\dot{M}_{out} = 7 \times 10^{-9}$ M$_{\odot}$ yr$^{-1}$ for $\epsilon_B = 0.01$ (primarily constrained by the luminosity of the radio light curve on the decline; taking $v_w = 10$ km s$^{-1}$). Figures~\ref{fig:mod1} and \ref{fig:mod2} show the synchrotron models superimposed on the observed light curves for $\epsilon_B = 0.1$ and $\epsilon_B = 0.01$, respectively. The models are similar between the two figures, although the transition from the ejecta-dominated phase to the Sedov phase occurs later in Figure~\ref{fig:mod1}, where $\dot{M}_{out}$ is smaller.
The model light curves peak significantly earlier and brighter than the observed light curve peaks, which can be reconciled because, at these early times, the shock is interacting with and being absorbed by the relatively dense $\dot{M}_{in}$ (so this bright emission is being obscured). By the time the shocks breaks out of $\dot{M}_{in}$ around day 25 and begins interacting with the less dense $\dot{M}_{out}$, the light curve models in Figure~\ref{fig:mod1} are declining from peak.  
For the models on the decline from light curve peak, the lower frequencies are brighter than the higher frequencies, as expected for our model selection of $p=2.1$ (the shallowest particle spectrum accommodated in the \citealt{Chevalier82, Chevalier98} model),  which yields $\alpha = -0.55$ (significantly steeper than observed in V745 Sco; \S \ref{sec:spec}). In estimating $\dot{M}_{out}$, we fit the synchrotron model to the data at an intermediate frequency, 4.6 GHz, in an attempt to capture the average behaviour of the multi-frequency light curve.

The optically thin evolution of the radio flux density (valid after radio light curve peak) scales as:
\begin{equation}\label{eq:thinflux}
S_{\nu} \propto \frac{R^3}{D^2} \epsilon_e\, \epsilon_B^{(p+1)/4} \rho_{CSM}^{(p+5)/4}\, v_s^{(p+5)/2}\, \nu^{(1-p)/2}
\end{equation}
At late times, the blast wave should be in its Sedov Taylor phase, and assuming it is interacting with a $\rho_{CSM} \propto r^{-2}$ wind-like CSM, the evolution of the blast-wave radius scales as $R \propto KE^{1/3} (\dot{M}/v_w)^{-1/3} t^{2/3}$ and the blastwave velocity scales as $v_s \propto KE^{1/3} (\dot{M}/v_w)^{-1/3} t^{-1/3}$\citep{Tang&Chevalier17}. Substituting into Equation \ref{eq:thinflux}, we find: 
\begin{equation}
S_{\nu} \propto \frac{KE}{D^2} \epsilon_e \epsilon_B^{(p+1)/4} \left(\frac{\dot{M}}{v_w}\right)^{(p+1)/4} t^{-(p+1)/2}\, \nu^{(1-p)/2}
\end{equation}
The implication is that our estimate of the CSM density depends on assumptions about the shock microphysical parameters as $\dot{M}/v_w \propto \epsilon_e^{-4/(p+1)} \epsilon_B^{-1}$. Therefore, for example, if $\epsilon_B \approx 0.001$ and $\epsilon_e = 0.004$ as modelled for radio supernova remnants \citep{Sarbadhicary+17} (rather than the $\epsilon_e = 0.1$ and $\epsilon_B = 0.01/0.1$ assumed above), the $\dot{M}/v_w$ we would infer from the synchrotron luminosity would increase to $\dot{M}_{out} = 5 \times 10^{-6}$ M$_{\odot}$ yr$^{-1}$. However, such a high $\dot{M}_{out}$ would imply a higher opacity to free-free absorption (Equation \ref{eq:tauff}) and a late light curve peak which is inconsistent with our observations  (especially at the lower frequencies). Therefore, the combination of constraints on synchrotron luminosity and $\tau_{ff}$ imply relatively efficient microphysical parameters governing particle acceleration and $B$ field amplification, $\epsilon_e \approx 0.1$ and $\epsilon_B \approx 0.01-0.1$, in V745 Sco.

\subsection{V745 Sco in Context: Symbiotic Stars and SN Ia Progenitors} \label{sec:snia}

As described in \S \ref{sec:model}, V745 Sco shows dense, absorbing CSM out to $\sim$ several $\times 10^{14}$ cm, and much lower density CSM at larger radii. It is not currently clear if the dense, close-in CSM is characterized by (a) a disc-like equatorial density enhancement \citep{Orlando+17}, which is presumably the product of wind Roche Lobe overflow and the mass transfer being concentrated in the orbital plane \citep{Mohamed&Podsiadlowski12},
or (b) a more spherical wind that has only had 25 years to fill the environment (the time elapsed between the 1989 and 2014 eruptions).  
The low-density wind at larger radii has $\dot{M}_{out} = 9 \times 10^{-10} - 7\times 10^{-9}$ M$_{\odot}$ yr$^{-1}$ for $v_{w} = 10$ km s$^{-1}$. This is significantly lower than the $\dot{M}$ values typically measured for symbiotic stars \citep{Seaquist&Taylor90, Seaquist+93}, but note that these works usually assume spherical, continuous, wind-like CSM. A relatively high accretion rate must be sustained on to the white dwarf in V745 Sco to drive the short nova recurrence time, and presumably this mass transfer is far from conservative, but much of the CSM resulting from non-conservative transfer is relatively hidden, tucked in close to the binary.

Is V745 Sco a viable progenitor system for Type Ia supernovae? Its relatively high mass and short recurrence time make it an intriguing candidate, but we can do one better: compare the properties of its CSM with constraints on CSM around SNe Ia, to test if a V745 Sco-like system has been ruled out as a progenitor of most SNe Ia. Constraints on CSM around SNe Ia come from radio observations \citep{Panagia+06, Chomiuk+16}, X-ray measurements \citep{Russell&Immler12}, and optical observations \citep[e.g.][]{Maguire+13}; see \citet{Maoz+14} for a review. Here we focus on the radio observations, because they are some of the most constraining and also are in direct analogy to the synchrotron-dominated radio light curve of V745 Sco presented here. Radio observations of SNe Ia are typically obtained $\sim 10-100$ days after SN explosion \citep{Chomiuk+16}, probing the CSM at radii of $\sim10^{16}-10^{17}$ cm through synchrotron emission. The nova blast in V745 Sco is $\sim$10 times slower than a SN Ia, indicating that the radius reached by the nova on day 20 will be reached by a SN Ia blast in just $\sim$2 days. The implication is that, if a SN Ia interacted with the dense, close-in CSM seen in V745 Sco ($\dot{M}_{in}$), this interaction would only be visible for a few days, and would have been missed by most extant radio observations of SNe Ia. 

Radio observations of SNe Ia would instead be probing larger radii, populated by the lower density, extended, wind-like CSM in V745 Sco ($\dot{M}_{out}$. Figure~9 of \citet{Chomiuk+16} shows that most radio observations of SNe Ia cannot constrain a wind below $\dot{M} <$ few $\times 10^{-9}$ M$_{\odot}$ yr$^{-1}$ for $v_{w} = 10$ km s$^{-1}$, $\epsilon_e = 0.1$, and $\epsilon_B = 0.1$, and $\dot{M} < 10^{-8}$ M$_{\odot}$ yr$^{-1}$ for similar parameters except $\epsilon_B = 0.01$. Assuming the $\epsilon$ microphysical parameters are similar in novae and SNe Ia, the implication is that most current radio observations of SNe Ia cannot rule out the CSM of V745 Sco, which is characterized by $\dot{M}_{out} = 9 \times 10^{-10}$ M$_{\odot}$ yr$^{-1}$ for $\epsilon_e = 0.1$, $\epsilon_B = 0.1$, and $\dot{M}_{out} = 7 \times 10^{-9}$ M$_{\odot}$ yr$^{-1}$ for $\epsilon_e = 0.1$, $\epsilon_B = 0.01$. Even simply using the constraint on $\dot{M}_{out}$ from $\tau_{ff}$ ($\dot{M}_{out} < 6 \times 10^{-8}$ M$_{\odot}$ yr$^{-1}$; Equation \ref{eq:tauff}), which does not depend on $\epsilon$ parameters, less than $\sim$half of SNe Ia with radio observations could rule out V745 Sco's CSM. 

We do note that a few of the nearest, best studied SNe Ia (e.g. SN\,2011fe, SN\,2014J) have unusually deep constraints on CSM \citep{Chomiuk+12b, Perez-Torres+14}, and these can rule out a V745 Sco like progenitor. We also note that perhaps more constraining than CSM observations is nebular spectroscopy of SNe Ia to search for material stripped from the companion star \citep[e.g.][]{Tucker+20}. These observations typically find $< 0.1$ M$_{\odot}$ of hydrogen can be swept up and remain consistent with spectroscopic limits, while $\ga$0.3 M$_{\odot}$ of material is expected to be stripped from a giant companion \citep{Pan+12, Boehner+17}. However, interpretation of nebular spectroscopy is model dependent \citep[e.g.][]{Botyanski+18}, and multiple constraints should be brought to bear in ruling out a SN Ia progenitor channel. In an era where investigators increasingly think that multiple progenitor channels contribute to the SN Ia population, and based on CSM properties alone, V745 Sco is allowed as a progenitor system for most SNe Ia.

\section{Conclusions}\label{sec:concl}
We have embarked on a multi-wavelength, but radio forward, investigation of the symbiotic recurrent nova V745 Sco. Our key results are:
\begin{itemize}
\item The line of sight to V745 Sco suffers dust extinction of $A_V = 2.14 \pm 0.6$ mag, as constrained by DIBs in optical spectroscopy, consistent with the total Galactic column along that line of sight (\S \ref{sec:dist}).
\item We estimate a distance of $8.2^{+1.2}_{-1.0}$ kpc to V745 Sco, constrained by its peak optical brightness, our extinction measurement, and a realistic three-dimensional map of the Milky Way and its dust (\S \ref{sec:dist}).
\item The apparent narrowing of optical emission lines during V745 Sco's 2014 eruption is likely not dominated by true deceleration due to interaction with CSM, but instead due to a rapid drop in line intensity and S/N. We therefore consider H$\alpha$ FWZI/2 measurements as lower limits on the shock velocity, except during the first observations, where we use FWZI/2 to measure the maximum ejecta velocity, 4450 km s$^{-1}$ (\S \ref{sec:decel}).
\item At radio wavelengths, V745 Sco is not detected during quiescence (2018--2023) by VLASS, consistent with a relatively low density CSM (\S \ref{sec:quiescence}).  
\item We find that the radio light curves of V745 Sco's 1989 and 2014 eruptions, constrained at $\sim$1.5 and $\sim$4.9 GHz,  are very similar between eruptions. However, the 1989 eruption is $\sim$20~per~cent fainter than the 2014 eruption, in contrast with the claim of \citealt{Kantharia+16} (\S \ref{sec:89v14}).  
\item The radio light curve during eruption is dominated by synchrotron emission, as determined from the brightness temperature evolution (\S \ref{sec:temp}). 
\item We apply a simple \citet{Chevalier82} model to the 0.6--37 GHz radio synchrotron light curves of V745 Sco in eruption, but find that interaction with a $\rho \propto r^{-2}$ wind-like CSM cannot explain the early portion of the light curves (on the rise to radio peak; the first $\sim$30 days of eruption). Instead, we find relatively dense CSM close-in to the binary ($\dot{M}_{in} = [5-10] \times 10^{-7}$ M$_{\odot}$ yr$^{-1}$ at $<$ several $\times 10^{14}$ cm, as primarily constrained by X-ray measurements; \citealt{Delgado&Hernanz19}), and a less dense, extended wind-like CSM at larger radii. The dense close-in CSM may be an equatorial density enhancement \citep{Orlando+17}, or it may be the red giant wind which has only had 25 years since the last nova eruption in which to pollute its environment \citep{Delgado&Hernanz19}.
We find that the lower density, outer wind can be characterized by $\dot{M}_{out} = 9 \times 10^{-10} - 7 \times 10^{-9}$ M$_{\odot}$ yr$^{-1}$, assuming $v_{w} = 10$ km s$^{-1}$, $\epsilon_e = 0.1$, and $\epsilon_B = 0.01-0.1$. We favor  efficient $\epsilon$ values such as these, because relatively high values are necessary to reconcile the luminous late-time synchrotron emission and the early radio light curve peak in V745 Sco (\S \ref{sec:model}). 
\item It would be relatively easy to hide a V745 Sco-like CSM around a SN Ia, because the dense CSM is confined to near the binary (which would only be constrained by early observations), and the extended wind is lower density than most constraints on SN Ia CSM. This implies that, based on CSM considerations alone, V745 Sco (and symbiotic recurrent novae like it) remains a viable SN Ia progenitor for a significant fraction of SNe Ia (\S \ref{sec:snia}).
\end{itemize}

In the future, we are excited to extend the simulations of \citet{Orlando+17} to simulate the multi-frequency radio light curve predicted for a nova interacting with a complex CSM with an equatorial density enhancement. This will give us a more intricate picture of the blast wave evolution that we can then compare with radio observations, and further elucidate the nature of the dense CSM close in to the binary.

\section*{Acknowledgements}
IM, LC, EA, KVS, PC, and CEH are grateful for support from NSF grants AST-1751874, AST-2107070, and AST-2205628, and NASA grants 80NSSC23K0497 and 80NSSC23K1247. JS acknowledges support from the Packard Foundation. EA gratefully acknowledges support from the NASA Hubble Fellowship. JLS acknowledges support from NSF grant AST-1816100.

The National Radio Astronomy Observatory is a facility of the National Science Foundation operated under cooperative agreement by Associated Universities, Inc. We acknowledge with thanks the variable star observations from the AAVSO International Database contributed by observers worldwide and used in this research. This work has made use of data from the European Space Agency (ESA) mission {\it Gaia} (\url{https://www.cosmos.esa.int/gaia}), processed by the {\it Gaia} Data Processing and Analysis Consortium (DPAC; \url{https://www.cosmos.esa.int/web/gaia/dpac/consortium}). Funding for the DPAC has been provided by national institutions, in particular the institutions participating in the {\it Gaia} Multilateral Agreement. 
This research has made use of the CIRADA cutout service at \url{cutouts.cirada.ca}, operated by the Canadian Initiative for Radio Astronomy Data Analysis (CIRADA). CIRADA is funded by a grant from the Canada Foundation for Innovation 2017 Innovation Fund (Project 35999), as well as by the Provinces of Ontario, British Columbia, Alberta, Manitoba and Quebec, in collaboration with the National Research Council of Canada, the US National Radio Astronomy Observatory and Australia's Commonwealth Scientific and Industrial Research Organisation.

\section*{Data Availability}
The optical spectra are publicly available at https://www.astro.sunysb.edu/fwalter/SMARTS/NovaAtlas/. All VLA data are publicly available from the NRAO data archive at data.nrao.edu under project codes AL202,AH383, AH389,220 and AH390 (for the 1989 observations) and  13B-057 (for the 2014 observations).

\bibliography{radionovae}

\begin{table}
\begin{center}
\caption{Spectroscopic Observations of H$\alpha$ Over the Course of V745 Sco's 2014 Eruption}
\label{tab:spectral}
\begin{tabular}{ccccc}
\hline\hline
UT Date & $t-t_0$ & FWHM & FWZI & Integrated Flux\\
 & (days) & (km s$^{-1}$) & (km s$^{-1}$) & (erg s$^{-1}$ cm$^{-2}$\AA$^{-1}$) \\
\hline
2014 Feb 09& 3 & 2233 & 8909.3 & 1.34e-08\\ 
2014 Feb 10& 4 &1484 & 8580.9 &1.12e-08\\
2014 Feb 11& 5 &1147 & 8293.6 & 6.51e-09\\
2014 Feb 13& 7 &889.3 & 8071.2 & 6.41e-09\\
2014 Feb 15& 9 &739.1 & 7335.4 & 3.37e-09\\
2014 Feb 16& 10 &682.7 & 5272.0 & 2.83e-09\\
2014 Feb 19& 13 &552 & 5452.4 & 1.05e-09\\
2014 Feb 20& 14 &509.9 & 4794.1 & 8.28e-10\\
2014 Feb 21& 15 &476 & 5084.8 & 6.44e-10\\
2014 Feb 22& 16 &455.5 & 4979.0 & 5.49e-10\\
2014 Feb 23& 17 &439.3 & 4467.3 & 4.74e-10\\
2014 Feb 24& 18 &433.3 & 4513.0 & 4.21e-10 \\
2014 Feb 25& 19 &423.2 & 4416.8 & 3.51e-10\\
2014 Feb 26& 20 &424.2 & 4237.7& 3.32e-10\\
2014 Feb 27& 21 &418.4 & 3995.7& 3.13e-10\\
2014 Mar 03& 25 &397.2 & 3385.6& 2.30e-10\\
2014 Mar 05& 27 &387.3 & 3344.6& 2.03e-10\\
2014 Mar 09& 31  &387 & 3051.6 & 1.33e-10\\
2014 Mar 10&  32 &380.3 & 2778.0 & 1.15e-10\\
2014 Mar 11&  33 &388.6 & 2583.7& 1.02e-10\\
2014 Apr 08& 61 &375.8 & 1833.6\\
2014 Apr 09& 62 &387.7 & 1797.1& 1.20e-11\\
2014 Apr 10& 63&343.7 & 1738.8 & 5.64e-12\\
2014 Apr 13& 66 &341.1 & 1706.0\\
2014 Apr 18& 71 &346.5 & 1643.5 \\
2014 Apr 22& 75&306 & 1613.5\\
2014 Apr 23& 76 &325.6 & 1571.4\\
2014 Apr 24& 77&253.7 &1507.9 & 1.64e-11\\
2014 Apr 27& 80 &362.3 & 1490.6 \\
2014 Apr 28& 81&323.9 & 1434.6\\
2014 Apr 30& 83&330 & 1410.5\\
2014 May 02& 85&362.8 & 1366.5\\
2014 May 04&87 &366.7 & 1355.5 \\
2014 May 05& 88&380 & 1186.5 & 4.42e-11\\
2014 May 06& 89&366 &  1190.3\\
2014 May 07&90 &366.9 & 1157.8\\
2014 May 08& 91&335.1 & 1139.1& 5.82e-12\\
2014 May 11& 94&332.6 & 1095.5\\
2014 May 12& 95&327.1 & 1080.3 & 1.92e-12\\
2014 May 13&96 &319 &  1072.0\\
\hline
\end{tabular}
\tablenotetext{a}{We take the start of eruption $t_0$ to be February 6, 2014. The integrated flux was measured after flux calibration.} 
\begin{flushleft}
\end{flushleft}
\end{center}
\end{table}

\begin{landscape}
\begin{table*}
    \centering
    \caption{\label{tab:2014} Radio Observations of V745 Sco's 2014 Eruption}
\hspace{-2in}
    \begin{tabular}{|l|c|c|c|c|c|c|c|c|c|c|}
\hline
UT Date& $t-t_0^{a}$ & \multicolumn{8}{c}{Flux Density $\pm$ 1$\sigma$ Error}   & Config\\
 &  & 1.25 GHz & 1.75 GHz  & 4.6 GHz   & 7.4 GHz   & 13.6 GHz  & 17.5 GHz   & 28.2 GHz  & 36.5 GHz  & \\
 & (days) & (mJy)&  (mJy) & (mJy)& (mJy)  & (mJy) & (mJy)  & (mJy) & (mJy) & \\
 \hline
2014 Feb 8.6& 2.6 & & & 0.13 $\pm$ 0.01 & 0.24 $\pm$ 0.02 & & & & & BnA\\
2014 Feb 9.6& 3.6 & & & 0.28 $\pm$ 0.02 & 0.40 $\pm$ 0.02 & & & 2.40 $\pm$ 0.24 & 2.50 $\pm$ 0.25 & BnA \\
2014 Feb 23.5& 17.5 & & &  6.80 $\pm$ 0.34 & 7.00 $\pm$ 0.35 & & &  7.70 $\pm$ 0.81 & 5.40 $\pm$ 1.14 & A \\  
2014 Mar 3.6& 25.6 & 7.33 $\pm$ 0.47 & 8.35 $\pm$ 0.44 & 8.93 $\pm$ 0.45 & 8.26 $\pm$ 0.41 &  7.92 $\pm$ 0.80 & 7.42 $\pm$ 0.75 & 5.90 $\pm$ 0.61  & 5.08 $\pm$ 0.54  & A\\ 
2014 Mar 16.5 & 38.5 & 4.85 $\pm$ 0.25 & 5.51 $\pm$ 0.28 & 6.00 $\pm$ 0.30 & 5.34 $\pm$ 0.27 & 4.92 $\pm$ 0.49 & 4.37 $\pm$ 0.44 & 3.12 $\pm$ 0.34 & 3.03 $\pm$ 0.43 & A \\  
2014 Mar 25.5 &47.5& 4.17 $\pm$ 0.21 & 4.59 $\pm$ 0.24 & 4.43 $\pm$ 0.22 & 4.01 $\pm$ 0.20 & 3.80 $\pm$ 0.38 & 3.62 $\pm$ 0.36 & 3.56 $\pm$ 0.36 & 3.22 $\pm$ 0.33 & A\\  
2014 Mar 31.5& 54.5 & 4.04 $\pm$ 0.21 & 4.40 $\pm$ 0.28 & 4.43 $\pm$ 0.22 & 4.00 $\pm$ 0.20 & 3.81 $\pm$ 0.38  & 3.64 $\pm$ 0.36 & 3.55 $\pm$ 0.36 & 3.77 $\pm$ 0.53 & A\\  
2014 Apr 9.4& 63.4 & 2.91 $\pm$ 0.18 & 2.87 $\pm$ 0.16 & 2.73 $\pm$ 0.14 & 2.56 $\pm$ 0.13 & 2.53 $\pm$ 0.26 & 2.61 $\pm$ 0.27 & 2.83 $\pm$ 0.30  & 2.90 $\pm$ 0.34 & A\\  
2014 Apr 11.5& 65.5 & 3.24 $\pm$ 0.18 & 3.13 $\pm$ 0.17 & 2.68 $\pm$ 0.14 &  2.41 $\pm$ 0.12 & 2.48 $\pm$ 0.25 & 2.50 $\pm$ 0.25 & 2.71 $\pm$ 0.28 &  2.86 $\pm$ 0.30 & A\\
2014 Apr 20.4& 74.4 & 2.78 $\pm$ 0.16 & 2.60 $\pm$ 0.14 & 2.21 $\pm$ 0.11 &  1.96 $\pm$ 0.10 & 1.99 $\pm$ 0.20 & 2.00 $\pm$ 0.20 & 2.42 $\pm$ 0.25 &  2.14 $\pm$ 0.24 & A\\
 2014 May 9.3& 92.3 & 1.85 $\pm$ 0.10 & 1.75 $\pm$ 0.10 & 1.43 $\pm$ 0.07 &  1.29 $\pm$ 0.07 & 1.38 $\pm$ 0.14 & 1.38 $\pm$ 0.14 & 0.91 $\pm$ 0.10 &  0.59 $\pm$ 0.08  & A\\
2014 Jun 1.3 & 116.3 & 1.82 $\pm$ 0.10 & 1.70 $\pm$ 0.10 & 1.05 $\pm$ 0.07 &  0.97 $\pm$ 0.05 & 1.08 $\pm$ 0.11 & 1.06 $\pm$ 0.11 & 1.02 $\pm$ 0.11  & 0.93 $\pm$ 0.11 & A\\
2014 Jul 3.5& 148.5 & $ <2.28$ & $<1.14$ & 0.76 $\pm$ 0.06  & 0.79 $\pm$ 0.04 & 0.96 $\pm$ 0.10 & 1.00 $\pm$ 0.11  & 1.14 $\pm$ 0.14 & 1.00 $\pm$ 0.15 & D\\
2014 Aug 7.1& 183.1 & $<1.78$ & $<1.36$ & 0.68 $\pm$ 0.05 & 0.61 $\pm$ 0.04 & 0.78 $\pm$ 0.09 & 0.70 $\pm$ 0.08 & 0.56 $\pm$ 0.13 & $<0.44$ & D\\
2014 Sep 29.9 & 236.9 & $<0.95$ & $<0.58$ & 0.49 $\pm$ 0.04 & 0.40 $\pm$ 0.03 & & & & &  DnC\\    
2014 Oct 8.0& 245 & & & & & 0.45 $\pm$ 0.05 & 0.45 $\pm$ 0.05 & 0.49 $\pm$ 0.08 & 0.32 $\pm$ 0.10 & DnC$\rightarrow$C$^b$\\
2015 Jan 30.6 & 359.6 & $<0.42$ & $<0.26$ & 0.27 $\pm$ 0.03 & 0.23 $\pm$ 0.02 & & & & & CnB$\rightarrow$B$^b$\\
2015 Feb 1.7 & 361.7 & & & & & 0.21 $\pm$ 0.02 & 0.18 $\pm$ 0.02 & 0.21 $\pm$ 0.04 & $<0.17$ & CnB$\rightarrow$B$^b$\\ 
\hline
\multicolumn{11}{l}{$^a$We take $t_0$ to be the time of discovery to be February 6.69, 2014.} \\
\multicolumn{11}{l}{$^b$Observations were obtained during `move time' between VLA array configurations.} \\
\end{tabular}
\end{table*}
\end{landscape}

\appendix
\section{Is the shock in V745 Sco radiative due to efficient particle acceleration?} \label{sec:app}

Here we further explore the question proposed in \S \ref{sec:decel}: is the rapid decline in H$\alpha$ line widths (Figure~\ref{fig:VFWHM}) merely an observational effect driven by dropping densities and fading lines, or does it imply dramatic energy loss from the shock due to acceleration of relativistic particles, as suggested by \citet{Delgado&Hernanz19}?

If the blast wave is losing energy due to efficient particle acceleration, this population of relativistic particles may have observable signatures in the form of GeV $\gamma$-rays. Marginal $\gamma$-ray detections had 2.5 and 2.4$\sigma$ significance on 2014 Feb 6 (the day of eruption discovery) and Feb 7 (day 1), with $>$100 MeV fluxes of $1.8\times10^{-7}$ phot cm$^{-2}$ s$^{-1}$ \citep{Cheung+14, Franckowiak+18}. However, these marginal detections occur before the steep drop in FWZI, which begins around day 10 (Figure~\ref{fig:Ha}). A more relevant constraint may be the sliding time window analysis of \citet{Franckowiak+18}, which found a 95~per~cent upper limit on the $>$100 MeV flux of $<0.5 \times 10^{-7}$ phot cm$^{-2}$ s$^{-1}$ over a time window covering 3--18 days following eruption. Assuming the same $\gamma$-ray spectrum as V906 Car (the highest S/N {\it Fermi}/LAT observation of a nova; \citealt{Aydi+20}) and a distance of 8.2 kpc, this translates to a $\gamma$-ray luminosity of $< 5.2 \times 10^{35} $ erg s$^{-1}$ and a $\gamma$-ray fluence over this time period of $<6.7 \times 10^{41}$ erg. Assuming that 20--30~per~cent of the energy in relativistic particles is radiated as GeV $\gamma$-rays \citep{Metzger+15}, this implies that $< 3.3 \times 10^{42}$ erg is transferred to relativistic particles during day 3--18. This is comparable to the kinetic energy of the ejecta assuming $M_{ej} = 10^{-7}$ M$_{\odot}$ ($KE \approx 5 \times 10^{42}$ erg), but much less than the KE for $M_{ej} = 10^{-6}$ M$_{\odot}$ ($KE \approx 5 \times 10^{43}$ erg). The {\it Fermi}/LAT constraints therefore imply that the shock in V745 Sco could be radiative if the ejecta mass is very small (as suggested by \citealt{Page+15}), but not if $M_{ej}$ is closer to $10^{-6}$ M$_{\odot}$.

The temperature of the hot shocked X-ray gas observed by {\it Swift}/XRT over the course of V745 Sco's 2014 eruption shows a decline in temperature which tracks the narrowing of the H$\alpha$ FWZI \citep{Delgado&Hernanz19}, with relatively constant temperature for the first few days of eruption followed by a rapid decline. This does not need to imply deceleration of the blast (as suggested by Delgado \& Hernanz), but could instead reflect adiabatic expansion and cooling of the shocked gas, if the interaction with CSM significantly weakens.

\begin{figure}
    \centering
    \includegraphics[width=80mm]{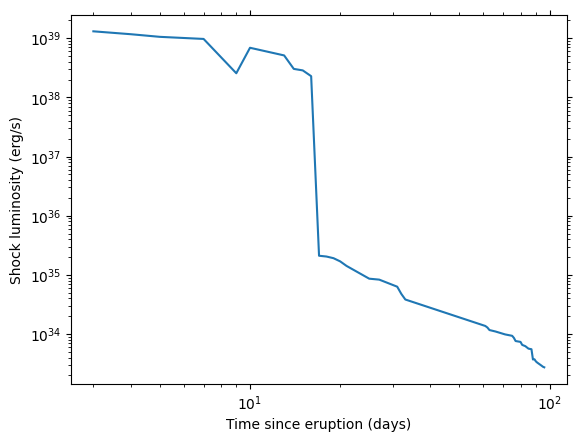}
    \caption{The estimated shock luminosity as a function of time over the course of V745 Sco's 2014 eruption. The sudden drop around day 16 is caused by the shock breaking out of relatively dense CSM into a much lower density environment.}
    \label{fig:shocklum}
\end{figure}

A radiative shock is not supported by the evolution of the X-ray absorbing column ($N_H$; Figure~4 of \citealt{Delgado&Hernanz19}), which shows that $N_H$ levels reach the ISM value just $\sim$16 days into eruption, around the time when the H$\alpha$ FWZI steeply declines. 
This rapid decline in $N_H$ and CSM density is further supported by our radio observations, which show all light curves peaking around this time near simultaneously---which must imply a sudden drop in the absorbing column (\S \ref{sec:model}). 
Both our radio observations and the analysis of \citealt{Delgado&Hernanz19} imply a dense CSM characterized by $\dot{M}_{in} \approx 5 \times 10^{-7}$ M$_{\odot}$ yr$^{-1}$ (assuming $v_w = 10$ km s$^{-1}$) out to a radius $\sim [1-9]\times 10^{14}$ cm, and a much lower density CSM characterized by $\dot{M}_{out} \la 10^{-9}$ M$_{\odot}$ yr$^{-1}$ at larger radii. 
If the shock was losing much of its energy to relativistic particles (as in the radiative shock scenario suggested by \citealt{Delgado&Hernanz19}), it would lose energy as a fraction of the shock luminosity, which can be written as  $L_{s} = 1.48 \times 10^{-24} R^2 n_{s} v_{s}^3$ \citep{Metzger+15}, where $n_{s}$ is the number density of particles in the shock (all quantities in cgs units). We estimate the shock luminosity as a function of time in Figure~\ref{fig:shocklum}, using $v_{s} = FWZI/2$, $R(t)$ as the integral under $v_{s}(t)$, and $n_{s}$ derived as $4\times$ the swept-up CSM density ($\dot{M}_{in}= 5 \times 10^{-7}$ M$_{\odot}$ yr$^{-1}$ and $\dot{M}_{out} = 10^{-9}$ M$_{\odot}$ yr$^{-1}$). Figure~\ref{fig:shocklum} shows that we expect the shock luminosity (and attendant energy loss through relativistic particles) to be much higher at early times, when the shock is interacting with denser CSM, and to plummet after day 16. 
 This is the opposite of what is required to explain the declining H$\alpha$ FWZI (Figure~\ref{fig:VFWHM}) through a radiative shock---the observed FWZI is quite constant during the first 10 days of eruption, and then rapidly declines between days 10--100. In other words---V745 Sco shows a temporal coincidence between the narrowing of the H$\alpha$ emission line profile and a sudden drop in the CSM density which is the opposite of what is expected if the shock was decelerating due to rapid particle acceleration. On the other hand, this temporal coincidence is expected if the narrowing of the lines is primarily an observational effect dictated by a drop in density (\S \ref{sec:decel}).

For these reasons, we do not treat the line width measurements as accurate tracers of the shock velocity, and instead employ a model for an adiabatic shock \citep{Tang&Chevalier17} in modelling the shock kinematics and interpreting the radio light curve.

\bsp   
\label{lastpage}
\end{document}